\documentclass[A4paper,11pt]{article}
\pdfoutput=1 

\usepackage{jinstpub} 
\usepackage[utf8]{inputenc}
\usepackage{lineno}
\usepackage{textcomp}
\usepackage{xcolor}
\usepackage{graphicx}
\usepackage{subcaption}
\usepackage{upgreek}
\usepackage[LGRgreek]{mathastext}
\usepackage[toc,title,page]{appendix}
\usepackage{heppennames2}
\usepackage{soul}

\newcommand{\bi}{\begin{itemize}}
\newcommand{\ei}{\end{itemize}}

\newcommand{\intL}{$\lambda_{int}$}

\newcommand{\radL}{$X_{0}$}

\newcommand{\ftfp}{{\tt FTFP\_BERT\_EMN}}

\newcommand{\GEANTfour} {{\textsc{Geant4}}\xspace}
\newcommand{\cm}{\ensuremath{\,\text{cm}}\xspace}
\newcommand{\GeV}{\ensuremath{\,\text{Ge\hspace{-.08em}V}}\xspace}

\newcommand{\cmsorcid}[1]{\href{https://orcid.org/#1}{\hspace*{0.1em\raisebox{0.45ex}{\includegraphics[width=0.7em]{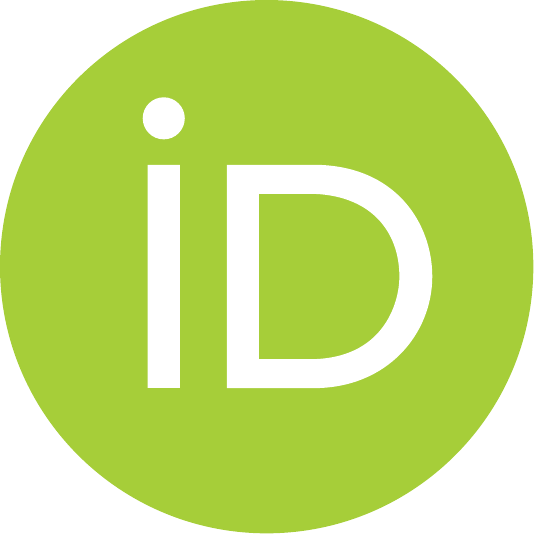}}}}}

\title{\boldmath Using graph neural networks to reconstruct charged pion showers in the CMS High Granularity Calorimeter}

\author[41]{M.~Aamir}
\author[60]{G.~Adamov\cmsorcid{0009-0007-1832-8038}}
\author[23]{T.~Adams\cmsorcid{0000-0001-8049-5143}}
\author[18]{C.~Adloff}
\author[67]{S.~Afanasiev\cmsorcid{0009-0006-8766-226X}}
\author[62]{C.~Agrawal}
\author[51]{C.~Agrawal}
\author[30]{A.~Ahmad\cmsorcid{0000-0002-4770-1897}}
\author[30]{H.~A.~Ahmed\cmsorcid{0000-0002-3558-0928}}
\author[30]{S.~Akbar}

% ...
\affiliation[18]{National Central University, \\ Chung-Li, Taiwan, ROC}
% ...
\affiliation[23]{Florida State University, \\ 600 W. College Ave., Tallahassee 32306, FL, USA}
% ...
\affiliation[30]{National Centre for Physics,Quaid-I-Azam University,\\ Islamabad-44000, Pakistan.}
% ...
\affiliation[41]{The University of Maryland,\\ College Park 20742, MD, USA}
% ...
\affiliation[51]{Indian Institute of Science Education and Research, \\ Dr. Homi Bhabha Road 411008, Pune, India}
% ...
\affiliation[60]{Georgian Technical University, \\Kostava str,  Tiiblisi, 77, 0160, Georgia.}
% ...
\affiliation[62]{Tata Inst. of Fundamental Research-B,\\Homi Bhabha Road, Mumbai 400005, India}
% ...
\affiliation[67]{Affiliated with an institute or an international laboratory covered by a cooperation agreement with CERN.}

\emailAdd{km.alpana@cern.ch}
\abstract{A novel method to reconstruct the energy of hadronic showers in the CMS High Granularity Calorimeter (HGCAL) is presented. The HGCAL is a sampling calorimeter with very fine transverse and longitudinal granularity. The active media are silicon sensors and scintillator tiles readout by SiPMs and the absorbers are a combination of lead and Cu/CuW in the electromagnetic section, and steel in the hadronic section. The shower reconstruction method is based on graph neural networks and it makes use of a dynamic reduction network architecture. It is shown that the algorithm is able to capture and mitigate the main effects that normally hinder the reconstruction of hadronic showers using classical reconstruction methods, by compensating for fluctuations in the multiplicity, energy, and spatial distributions of the shower's constituents. The performance of the algorithm is evaluated using test beam data collected in 2018 prototype of the CMS HGCAL accompanied by a section of the CALICE AHCAL prototype. The capability of the method to mitigate the impact of energy leakage from the calorimeter is also demonstrated.}

\begin{document}
\maketitle
\flushbottom

\newpage

\section{Introduction} \label{sec:intro}

Among other physics cases, the high-luminosity running of the LHC (HL-LHC) will provide the data required to measure the Higgs couplings to a precision of approximately 1\% and a first measurement of its triple self-coupling~\cite{bib.HLLHCYellowReport}. 
Operations of the HL-LHC will start towards the end of this decade with a planned instantaneous luminosity of 5 $\times$ 10$^{34}$ cm$^{-2}$s$^{-1}$ and an average number of interactions (pileup) per bunch crossing of 140 or more \cite{bib.cms-hgcalphase2-tdr}. 
The endcap calorimeters of the CMS experiment that cover a pseudorapidity range of $1.5<|\eta|<3.0$ will be replaced with two high granularity calorimeters (HGCAL) designed to withstand the radiation exposure expected in the ten or more years of the HL-LHC operations \cite{bib.cms-phase2-techproposal}.
In the technical design report \cite{bib.cms-hgcalphase2-tdr}, each HGCAL is a sampling calorimeter with a 28-layer electromagnetic section (CE-E) followed by a 22-layer hadronic section (CE-H). 
Recently the number of the CE-E and CE-H layers has been reduced to 26 and 21, respectively.
The active detectors in the CE-E and CE-H are hexagonal 8-inch silicon sensors subdivided into 0.56 and 1.26\,$\cm^2$ hexagonal cells, to be used in the regions of higher and lower radiation levels, respectively. 
In the regions away from beam pipe and deep into the calorimeter, where radiation levels are sufficiently low, plastic scintillator cells ranging in size 5 to 25\,$\cm^2$, which are directly read out by silicon photomultipliers are used.
In all, the total number of detector channels will be in excess of six million.
This fine segmentation has been chosen for efficient particle identification, particle-flow reconstruction, and pileup rejection.

In 2018 a prototype of the HGCAL calorimeter was exposed to the beam in the H2 beamline from the CERN SPS~\cite{bib.cern-h2-manual}.
The prototype consisted of a 28-layer electromagnetic section with 6-inch silicon sensors subdivided into $1.1\cm^2$ hexagonal detector channels and a 12-layer hadronic section, also equipped with silicon sensors.
Digitization of the signal in each channel was performed with the SKIROC2-CMS ASIC \cite{bib.Skiroc2cms}, which measured both amplitude and time-of-arrival of the signal.
The CALICE Analog Hadronic Calorimeter (AHCAL) \cite{bib.calice-ahcal-construction, bib.calice-ahcal-2018tb-paper} consisting of 39 layers of plastic scintillator tiles sandwiched between steel plates were placed at the back.
Together they made a reasonable approximation of the final HGCAL design.
To best approximate the sampling in the hadronic section of the HGCAL, only every fourth layer of the AHCAL was used in the analysis described here.
The prototype was exposed to beams of positrons and pions with momenta between 20 and 300\,\GeV, and muons with momenta of 200\,\GeV.
Details of the mechanical design or the readout electronics can be found in Refs.~\cite{bib.cern-h1paper,bib.cern-h2paper}.
The performance of the prototype calorimeter using positron and pion beams, evaluated with the conventional weighted summation method, has been previously reported in Refs.~\cite{bib.hgcal-2018-positrons} and~\cite{bib.hgcal-2018-pions}. These results have been used to benchmark and to demonstrate the accuracy of the \GEANTfour simulation~\cite{bib.GEANT4} of the prototype.

When hadrons are incident on a calorimeter detector, there are a wide variety of nuclear processes that take place as the energy is absorbed.
This leads to large event-by-event fluctuations in the types and multiplicity of shower particles, in the spatial distribution of their energy deposits, and in the fraction of invisible energy lost to nuclear binding energy, all of which depend on the energy of the incident particle.
The energy reconstruction of these showers can be further complicated by energy leakage due to insufficient detector coverage or energy losses due to missing channels.
Given the multidimensional nature of the problem of reconstructing the energy of hadronic showers, traditional algorithms such as the weighted summation method (WS method) described in~\cite{bib.hgcal-2018-pions}, are often complex, involve ad-hoc choices of parameterization, and do not fully mitigate the effects mentioned above.
On the other hand, graph neural networks (GNNs) are particularly well-suited for reconstructing showers in HGCAL, as the number of active cells in hadronic showers is typically a few hundred out of tens of thousands of cells in the region, and the spatial distributions of the active cells vary considerably from event to event~\cite{bib.gnns-particlereco}. 

In this publication, we present a novel algorithm to reconstruct the energy of charged pion showers with a GNN based dynamic reduction network (DRN) architecture~\cite{bib.drn-pointclouds}. 
A significant improvement in the energy resolution compared to the WS method is achieved using this algorithm.  
The improvements to the performance are validated using both the \GEANTfour simulation and the beam test data collected in 2018. 
We also show how the algorithm performs for a calorimeter with a geometry similar to the final HGCAL detector.

We describe briefly the experimental setup of the 2018 prototype beam tests in Section \ref{sec:tb2018}. The DRN model and procedure of energy reconstruction using the DRN are described in Section \ref{sec:mltools}. Performance comparisons with the Monte Carlo and with the beam test data are presented in Section \ref{sec:drn-reco} and \ref{sec:drn-tbdata-sim}, respectively. The performance of the algorithm when applied to simulated events in the HGCAL geometry is summarized in Section \ref{sec:fullhgcal}.

\section{The HGCAL Prototype, event reconstruction and simulation} \label{sec:tb2018}

\begin{figure}[htbp]
  \centering
  \includegraphics[width=0.75\linewidth]{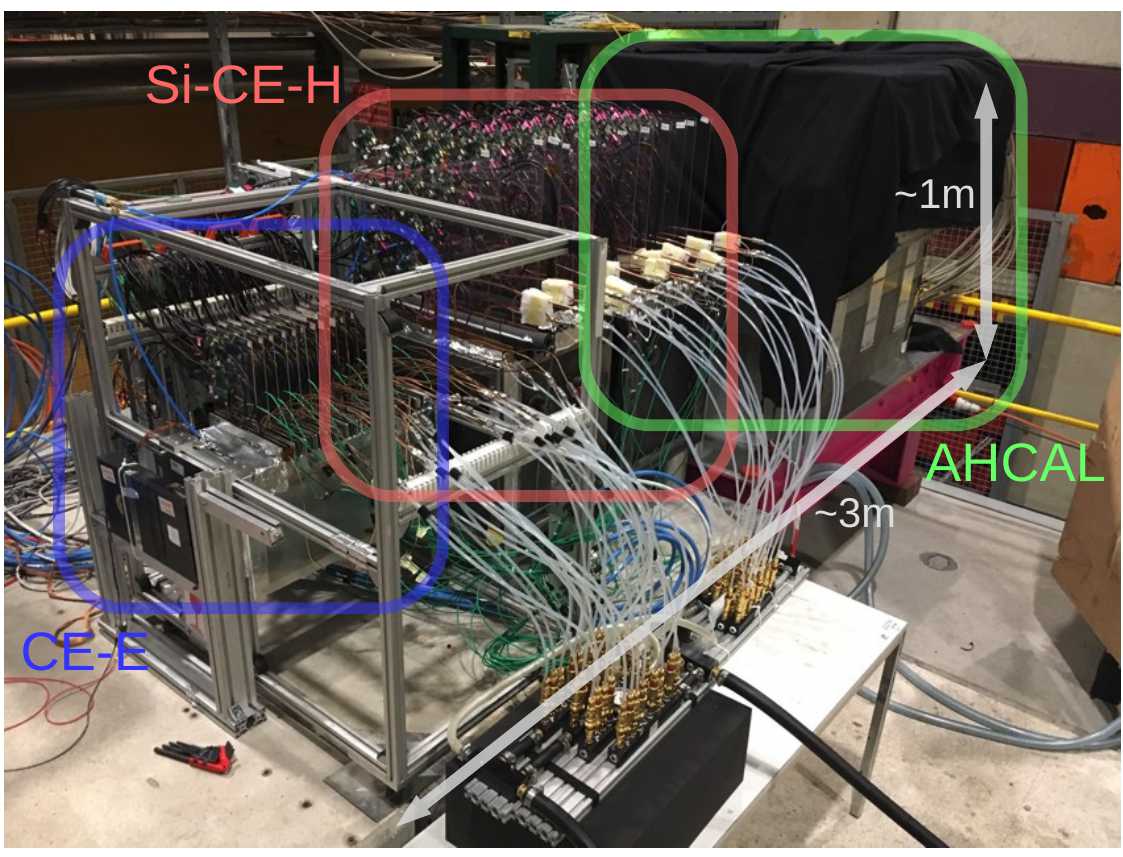}
  \caption{\label{fig:tb-hgcal-exp} A picture of the CE-E, CE-H, and AHCAL prototypes setup mounted on a concrete platform in the H2 experimental area of the CERN SPS.}
\end{figure}

The HGCAL prototype detector used in the beam test in 2018 consisted of a silicon sampling electromagnetic section (CE-E prototype) and a silicon sampling hadronic section (CE-H prototype). 
These were followed by the CALICE AHCAL prototype, which is a sampling calorimeter using scintillator tiles directly read out by SiPMs \cite{bib.calice-ahcal-2018tb-paper}. 
The experimental setup is described in \cite{bib.cern-h2paper} and is shown in Fig.~\ref{fig:tb-hgcal-exp}.
The basic detector unit of the CE-E and CE-H prototypes was a silicon module consisting of a 6-inch hexagonal silicon sensor and a printed circuit board (PCB) with the readout electronics, along with a baseplate for mechanical support. 
Each silicon sensor was subdivided into 128 hexagonal cells with an area of approximately 1.1 $cm^2$, and each cell was readout independently with signal amplitude and timing information being collected.

The CE-E prototype was longitudinally segmented into 28 sampling layers with one sensor module per layer. 
The absorber layers alternated between the layers made with a 6 mm thick copper cooling plate and a 1.2 mm thick CuW baseplate, and layers made with 4.9 mm thick Pb plates clad with 300 $\mu m$ thick stainless steel.
The total depth of CE-E was 26 \radL{} or 1.4 \intL. 
The CE-H prototype consisted of a stack of 12 layers of silicon modules, each mounted on a 6 mm thick copper cooling plate, sandwiched between 4 cm thick steel absorber plates. 
Each of the first 9 layers had seven silicon modules arranged in a daisy-like structure, while the last three layers were only instrumented with one module. 
The total depth of the CE-H prototype was approximately 3.4 \intL. There were $\sim$3500 channels in CE-E and $\sim$8500 channels in the CE-H prototype. 
The AHCAL section was made of 39 longitudinal sampling layers each composed of $3\times3\times0.3~cm^3$ scintillator tiles individually read out by SiPMs and steel absorber plates, and it added a further 4.4 \intL{} to the calorimeter stack. 
There were approximately 22,000 scintillator tiles, wrapped in reflective foils, read out independently.\\

For each event, the signal amplitude and timing information for each detector cell are saved as raw data for offline reconstruction of energies deposited and time of arrival of the signal~\cite{bib.hgcal-2018-pions}. 
The digitized signal from each readout channel, after subtracting pedestal noise, was converted into the corresponding energies measured in the unit of number of minimum ionizing particles (MIPs). 
The MIP conversion factor for each channel is obtained by fitting the digitized signal obtained with 200 GeV muons with a Landau distribution convoluted with a Gaussian distribution. 
The maxima of the fitted function is the MIP calibration value for that cell.

\begin{figure}[h]
  \centering
  \includegraphics[width=0.85\linewidth]{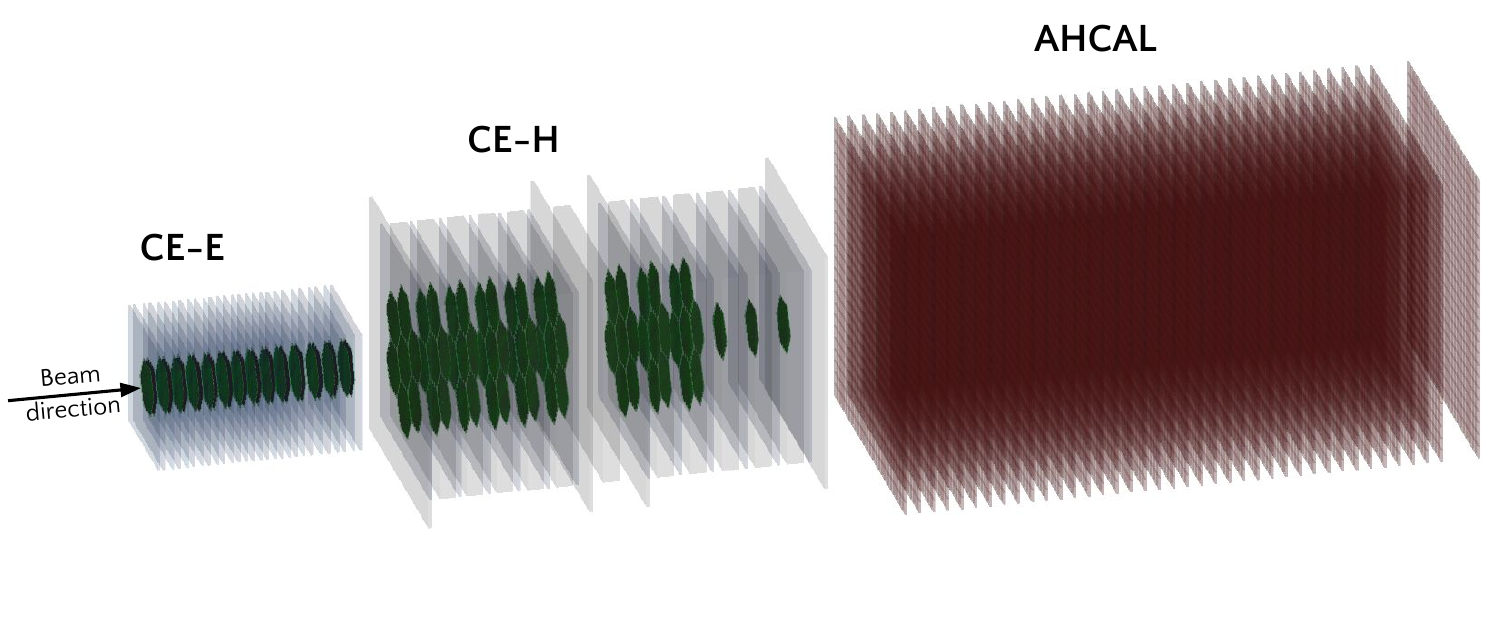}
  \caption{\label{fig:tb-hgcal-sim} Simulated geometry of the CE-E, CE-H, and AHCAL detector prototypes depicting the position of active layers and absorbing material.}
\end{figure}

The detector setup, along with the beamline elements, was simulated using a \GEANTfour simulation~\cite{bib.GEANT4}. 
The beamline elements, starting from the production target T2 up to the front of the HGCAL prototype, were simulated using the G4Beamline simulation framework implemented in \GEANTfour version 10.3. 
The calorimeter sections were simulated using \GEANTfour version 10.4.3. 
A schematic of the simulated HGCAL and AHCAL prototype detectors used in the beam test experiment is shown in Fig.~\ref{fig:tb-hgcal-sim}. 
The details are summarized in Ref.~\cite{bib.hgcal-2018-pions}. 
The simulated signal is also converted into units of the number of MIPs using the same procedure as used for the data.
The position of cells is defined in a right-handed Cartesian coordinate system with the z-axis parallel to the beam direction.

A sample of five million charged pions with a flat energy distribution between 10 and 350 GeV was produced using the \GEANTfour toolkit implemented in the CMS software framework with \ftfp{} physics list~\cite{bib.g4validation}. This sample is used to train the neural network models investigated for the energy reconstruction of hadron showers produced by charged pions. 

\section{Machine learning strategy and the dynamic reduction network} \label{sec:mltools}
The highly granular nature of the HGCAL makes it very suitable for the application of machine learning (ML) techniques to reconstruct particle showers in the calorimeter. 
Compared to other calorimeter designs, the level of detailed information in HGCAL is significantly higher. Machine learning techniques, which can process all or a substantial portion of this data, are particularly effective at leveraging these details to accurately reconstruct hadronic and electromagnetic showers.

These showers have large event-by-event variations in the spatial distributions of energy deposits in the detector cells and are usually interspersed with channels without a measurable signal.
Traditional ML architectures like multilayer perceptron neural networks (MLPs)~\cite{bib.tensorflow2015-whitepaper} and boosted decision trees (BDTs), expect fixed-size human-designed input features for reconstructing these showers, and it is difficult to design such features to efficiently capture these variations.
Modern ML algorithms, such as convolutional neural networks (CNNs), possess powerful feature-learning capabilities, and have demonstrated their ability to outperform traditional ML algorithms, which use human-designed input features, particularly when provided with low level detector information, like calorimeter hits~\cite{ImageNet,CNNNeutrino,DNNMicroBoone,CMSHtoAA,NNforhgcal-paper}.

While image based approaches like CNNs do use low level detector information, the requirement of fixed size inputs  necessitate zero-padding or truncation of the input collection. 
Additionally it is not possible to naturally represent information from multiple layers of detectors with varying granularity in square or rectangular images.
Finally, these architectures tend to impose ordering on their inputs that is inappropriate for this permutation-invariant collection of calorimeter hits. 
A natural way to represent these sparse, high-dimensional, variable-geometry sets of input features is with graphs. 
In particular, we represent each pion shower as a point cloud of hits in a high-dimensional latent space and then create graphs by drawing edges between nearby hits. 
Information is allowed to flow along the graph edges in GNNs as in CNNs, but with the advantage of, among others, flexibility in morphology. 

In electromagnetic showers, bremsstrahlung and pair-production processes dominate the energy loss until the shower components' energy falls below the critical energy, typically of the order of 10~MeV.
The fluctuations in the amount of energy that is undetected are small and do not affect significantly the energy resolution. 
However, in hadronic showers a large component of the energy is lost through excitation or disintegration of the absorber nuclei, and, since it is generally undetected, it is known as the `invisible energy' and can amount to up to 40\% of the total energy of a shower. 
Additionally, charged and neutral pions produced in nuclear interactions leave different energy depositions as neutral pions almost instantaneously decay to a pair of photons giving rise to electromagnetic showers while charged pions develop into hadronic showers.
Fluctuations in both the production rate of neutral or charged pions, and in the energy lost as `invisible energy' lead to significant event-by-event variations in the energy response of a calorimeter to incoming hadrons.   

In the conventional WS method used to determine the energy of an incoming hadron in a calorimeter that is divided into a few sampling layers (for example the 4 or 5 layers of the current CMS hadron calorimeter~\cite{bib.cms-phase-one-upgrade}), where the first layer is usually an electromagnetic calorimeter, the energy deposition in each layer is added with a different weight that is determined either empirically or by simulation. 
This method does not fully account for the large event-by-event fluctuations in hadronic showers and requires fine tuning for different regions of the detector, as well as for any dependence on the energy of incident hadrons.
This method works well for electromagnetic showers, which tend to be uniform, whereas for hadronic showers fluctuations limit its efficacy~\cite{bib.hgcal-em-optimization}.
In the context of high granularity calorimeters, the method has been extended by using local energy densities as a measure of the electromagnetic component of hadronic showers as described in~\cite{bib.calice-software-compensation}.

The HGCAL, with its fine longitudinal and transverse granularity, produces a highly detailed image of the longitudinal and lateral development of the shower. 
By using ML methods the information in this three-dimensional image can be used in a structured way to derive the energy and other properties of the incoming hadron. 
In addition, in the HGCAL, the arrival time of the signal in each readout channel will be available, adding another dimension to the shower reconstruction. 
However, the potential improvements expected by including timing information for event reconstruction have not been explored in this work. 

\begin{figure}[h]
    \centering
    \includegraphics[width=0.95\textwidth]{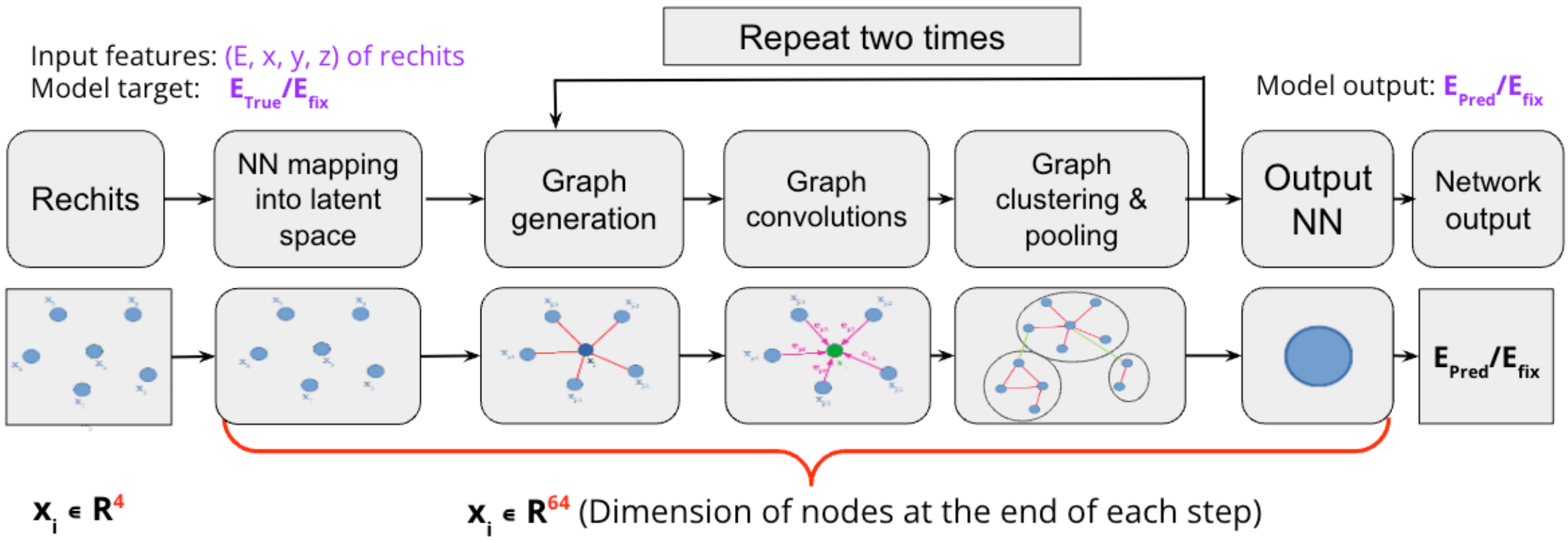}
    \caption{This schematic illustrates the GNN-based dynamic reduction network. The $x_i$ represents rechits, with red edges showing their connections in the latent space which are updated during graph convolution (in magenta). In the graph clustering stage, small circles indicate clusters, and nodes within as blue dots. Final predictions are produced after passing through the output neural network.
}
    \label{fig:drn}
\end{figure}

The DRN~\cite{bib.drn-pointclouds} builds upon dynamic graph convolution networks~\cite{bib.drn} with two modifications: firstly the graph operations are performed in a higher dimensional latent space, and secondly, there is an additional graph clustering and pooling step used to iteratively accumulate information across the graph. A schematic of such a network used for the studies presented in this paper is shown in Fig.~\ref{fig:drn}. A summary of the information flow from the inputs -- the spatial coordinates of the hits and their energy measured in each detector readout channel -- to the required target is as follows. 
\begin{enumerate}
\item The input features, attributes of detector channels such as spatial coordinates (x, y, and z) or energy (E), are normalized using fixed values such that most of the values lie in the range from zero to one, with outliers allowed. The normalized input data are mapped onto a 64-dimensional latent space using a fully connected neural network, which is an MLP encoder with a depth of three layers that are connected via an exponential linear unit activation function. 
\item High-level features are learned using GNN techniques by repeating the following three steps twice: 
\begin{enumerate}
\item Graph generation step: each point in the point cloud in the 64-dimensional latent space is connected to its 16 nearest neighboring nodes (including itself) to form graphs. 
\item Graph convolution step: the resulting graphs are processed using the "EdgeConv" operation~\cite{bib.edgeconv}, where two-layer fully connected networks are used to create "messages" that are sent along the edges of the graph. These "messages" are aggregated at each node in the graph to form the new node features.
\item Graph clustering and pooling: the graph convolution step is followed by graph clustering using the \textit{graclus} clustering algorithm~\cite{bib.graclus} and a pooling step, in which the updated features are weighted by distance and the clusters are combined using a max-pooling strategy to form a single point. The \textit{graclus} clustering algorithm successively and pairwise, combines connected graph nodes within a cluster that maximizes the resulting edge weight of the combined node. 
\end{enumerate}
Each pass through these three steps -- graph generation, graph convolutions, and graph clustering and pooling -- halves the number of nodes in the graph. 

\item The final reduced set of nodes are globally max-pooled and passed through a fully connected neural network, with an exponential linear unit activation function used at the encoder step, to obtain the final target, which in our case is the pion energy. 
\end{enumerate}
As described in Ref.~\cite{bib.drn-pointclouds}, the clustering and pooling steps are optimized in a way to maximize the reduction in information passed on for further processing, hence the network is called a "dynamic reduction network (DRN)". 

\section{Energy Reconstruction of pion showers using DRN method} \label{sec:drn-reco}

A sample of simulated hadron showers produced by negatively charged pions, uniformly sampled in the energy range from 10 to 350~GeV, is used to train the DRN for the reconstruction of the energy of incident pions. 
The hadron showers were simulated using the same \GEANTfour geometry and physics lists that were used for the simulation of events in the beam tests described above.
We apply the same selection criteria to remove events where the particles loose a large fraction of their energy in the beamline and apply the same noise rejection thresholds that were used in the data~\cite{bib.hgcal-2018-pions}.
The input to the DRN is a point cloud of the detector channels associated with a given hadronic shower. 
The attributes of each channel, used for training the model, are its position coordinates, $x$, $y$, and $z$, given in centimeters and the amplitude of the energy deposition in the sensor as either the number of MIPs or in terms of the effective energy in GeV, defined below.

As this is a sampling calorimeter, we need to infer from the signal in each sensor the energy deposited in the absorbers. 
In the prototype detector, the absorbers in CE-E, CE-H, and AHCAL were different and hence energy scale factors need to be estimated separately for them.  
We estimate the energy deposited in the absorber that corresponds to the observed signal amplitude using a MIP-to-GeV conversion factor. 
For CE-E this factor is estimated with 50 GeV positrons. In CE-H and AHCAL the MIP-to-GeV conversion factor is estimated with 50 GeV pions that do not initiate a hadronic shower in CE-E. 
We obtain factors of approximately 10.5 MeV per MIP and 80 MeV per MIP for the electromagnetic and hadronic sections, respectively.  

\begin{figure}[ht]
    \centering
    \includegraphics[width=0.45\textwidth]{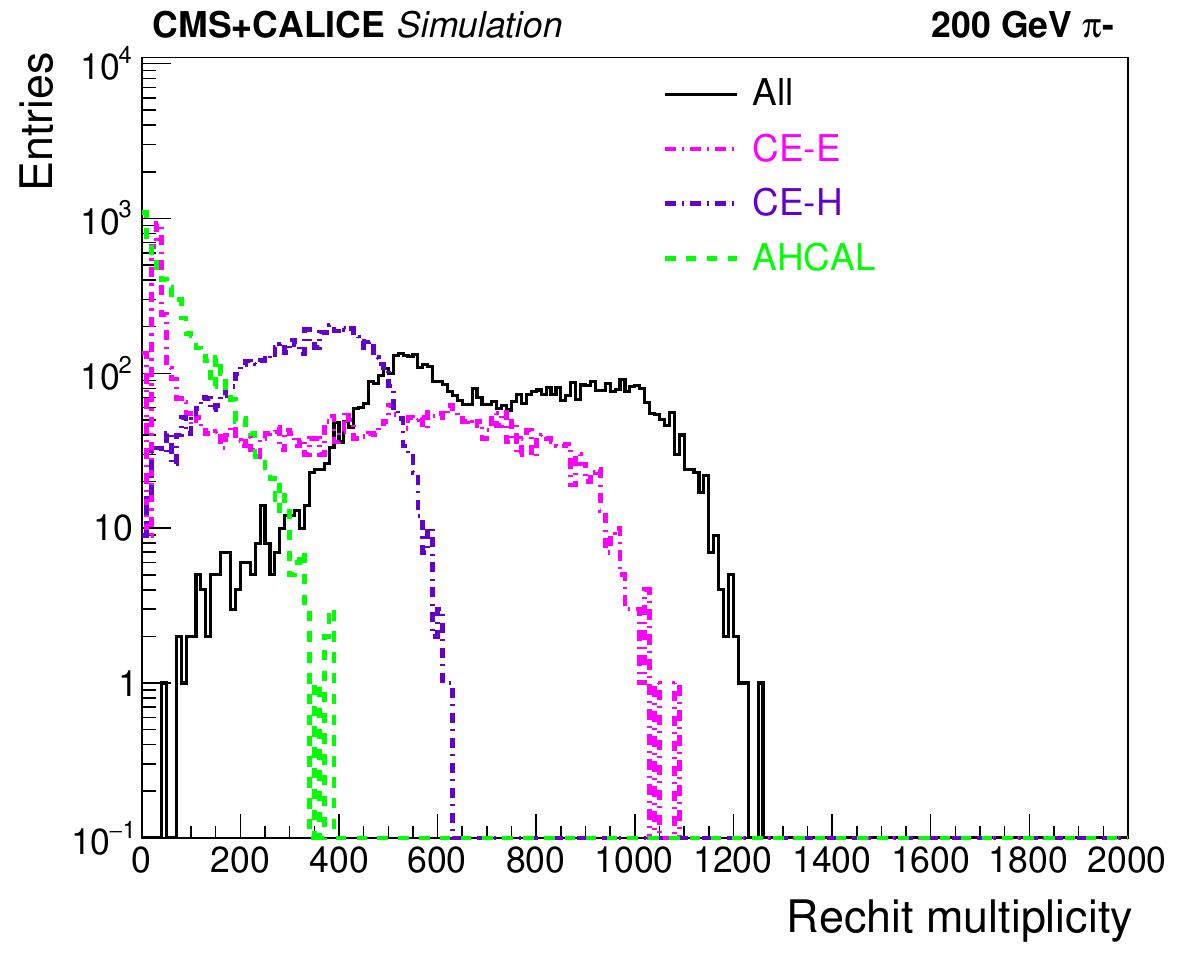}
    \includegraphics[width=0.45\textwidth]{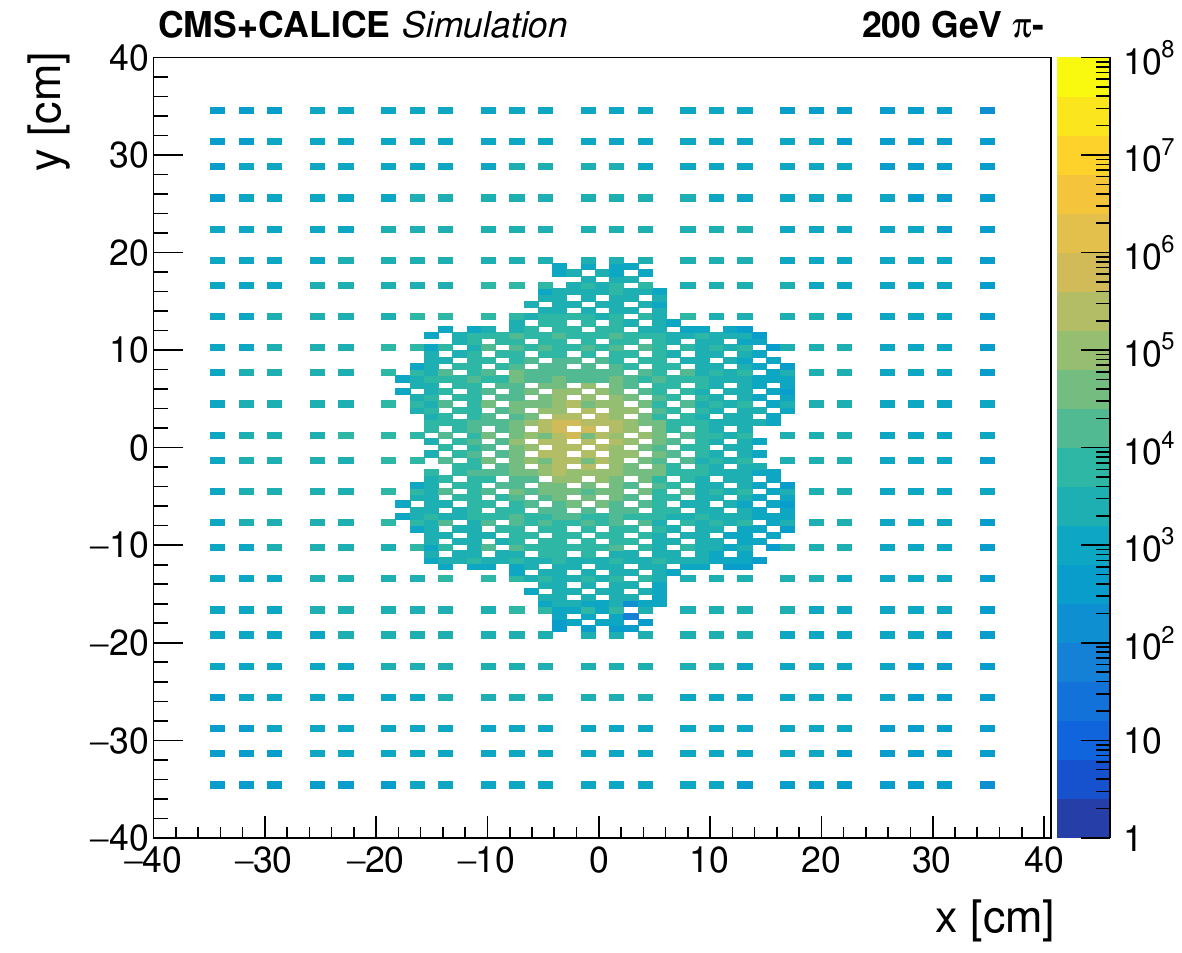}
    \includegraphics[width=0.45\textwidth]{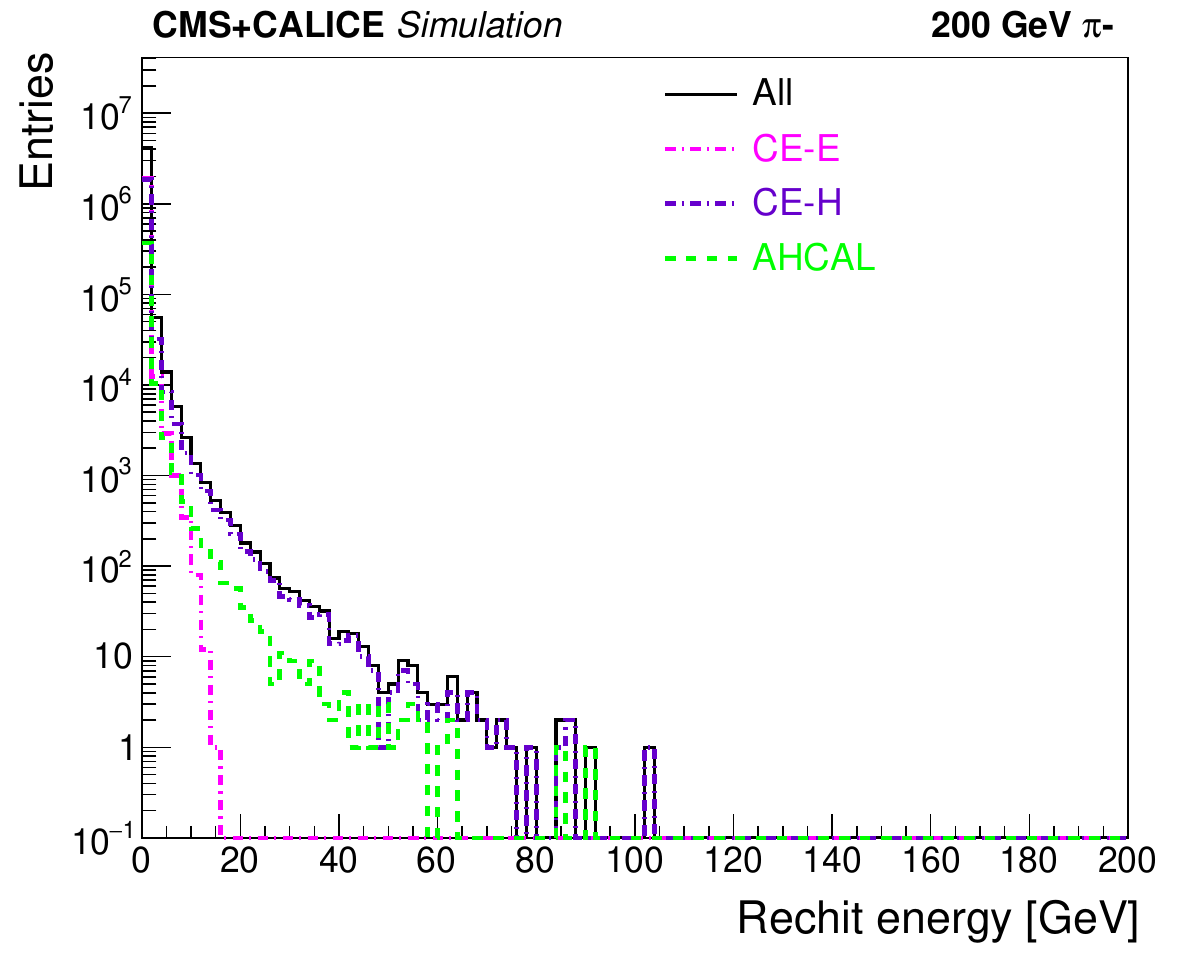}
    \includegraphics[width=0.45\textwidth]{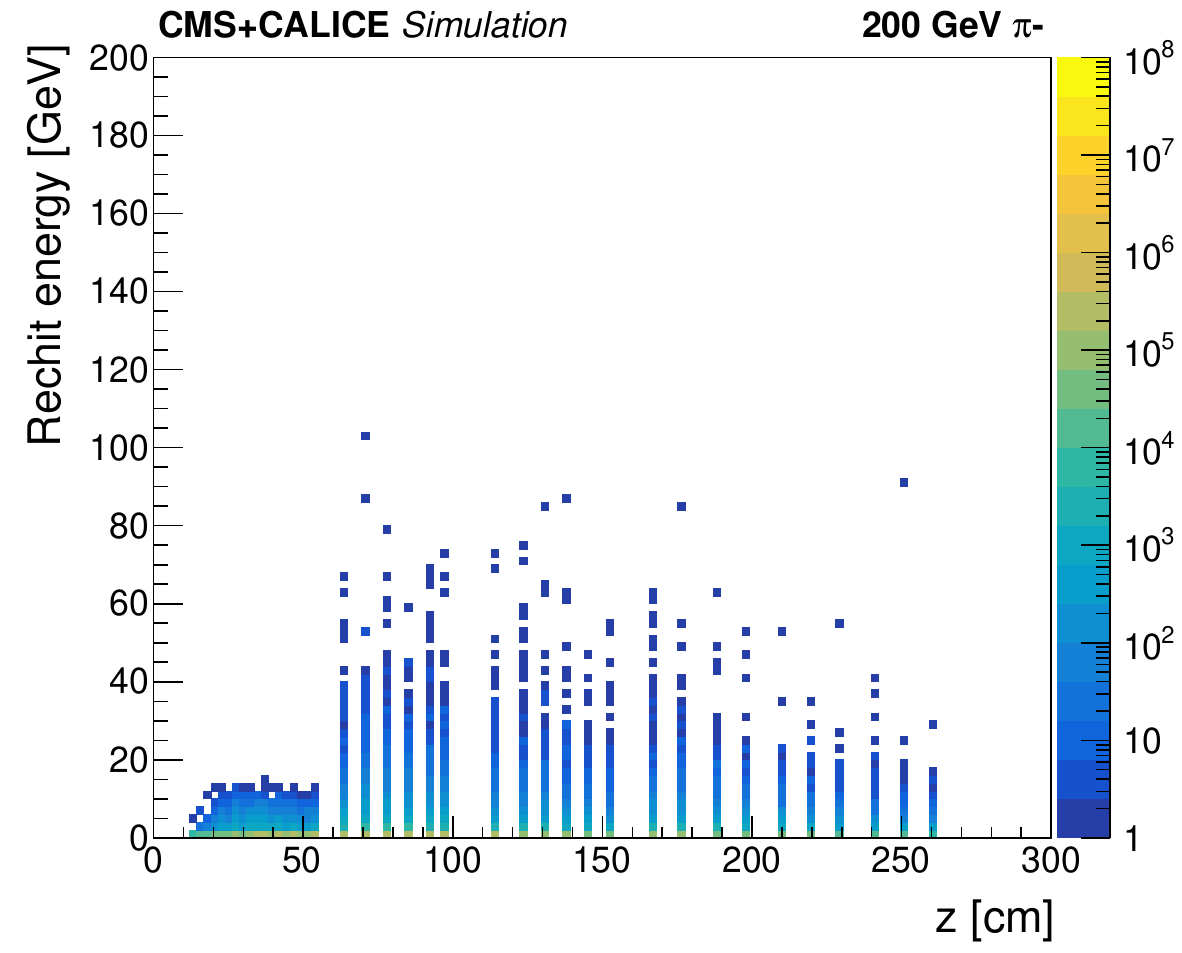}    
    \caption{A few input features used for training DRN, namely distributions of number of rechits (top left), $x$ and $y$ coordinates of rechits (top right), distribution of energy measured in individual rechits (bottom left), and distribution of rechit energies versus their depth in $z$ (bottom right).}
    \label{fig:input-features}
\end{figure}

The signals in the detector cells, weighted by their corresponding MIP-to-GeV factor, are used to estimate the amount of energy deposited in the absorber plates. 
These estimates of energy deposited in the absorber together with their positions are referred to as "reconstructed hits" or "rechits".
Distributions of the number of rechits with a signal measured above the noise thresholds in CE-E, CE-H, and AHCAL sections, and their $x$ and $y$ coordinates are shown in Fig.~\ref{fig:input-features} (top left) and Fig.~\ref{fig:input-features} (top right), respectively, for simulated 200 GeV $\pi^-$ data.
The distributions of the energy estimated in units of GeV in the
detector cells of the three detector sections are shown in Fig. \ref{fig:input-features} (bottom left).
Fig.~\ref{fig:input-features} (bottom right) shows the distribution of rechit energies along the depth of the detector (z coordinate).

The energies of all rechits can be summed to obtain an estimate of the total energy of the particles; this sum is referred to as $E_{fix}$ in the following.
However, since the response of a sampling calorimeter, such as the HGCAL prototype, to hadrons is not linear with the incident particle's energy, $E_{fix}$ is a suboptimal estimate of the particle's true energy. 
The WS method uses a $\chi^2-minimization$ method for a given incident particle energy, as discussed in~\cite{bib.hgcal-2018-pions}, to obtain weights to combine energies measured in CE-E, CE-H, and AHCAL sections.
Although the linearity of the response can be restored and the resolution improved, this method does not account for event-by-event fluctuations in the hadronic showers. 

We have trained a DRN model with simulated hadronic showers generated with a flat incident energy distribution.
The target in the ML training was the ratio of the true energy of the particle to that measured using the fixed MIP-to-GeV weights ($E_{fix}$). In the training, the learning rate was set to 0.0001. 
We trained three different models with three different input features, the energy ($E$), the energy and the longitudinal position ($E, z$), and the energy and both the lateral and the longitudinal positions ($E,x,y,z$) of each detector channel. 
These models are trained until the value of the loss function evaluated on the validation dataset stops improving in subsequent epochs. The epoch numbers 110, 114, and 135 correspond to the minimum loss values for DRN(E), DRN(E,z), and DRN(E,x,y,z), respectively, and corresponding training weights are used for evaluating the performance of pion energy reconstruction in the following.
Approximately 20\% of the generated events were reserved in a separate dataset for validation of the models using the network parameters derived in the training procedure. 
It took approximately 25 minutes to train a model for a single epoch on 4 NVIDIA V100 GPUs, with 192 GB RAM on each of the 4 GPU nodes, in parallel~\cite{bib.pytorch}. 

\begin{figure}[ht]
    \centering
    \includegraphics[width=0.32\textwidth]{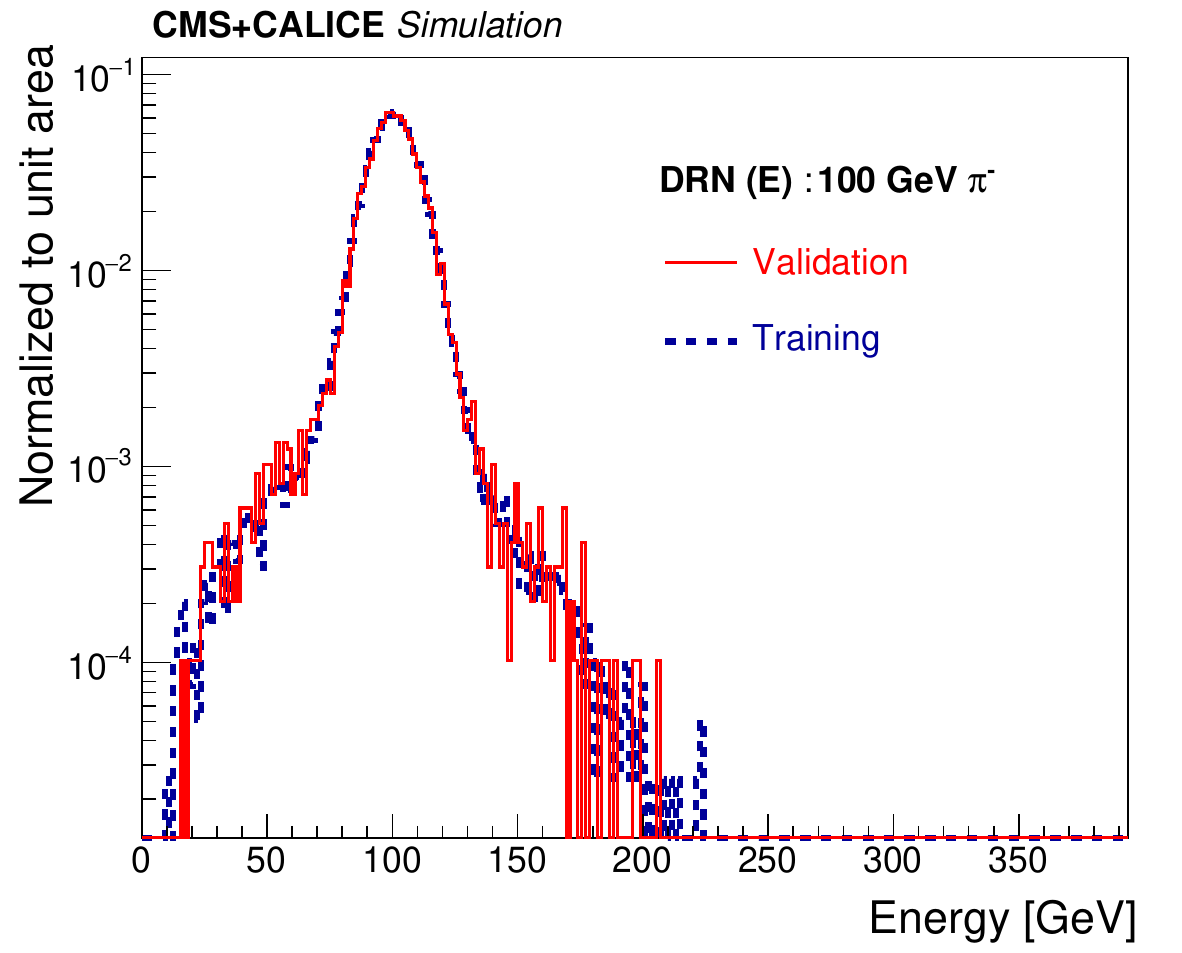}
    \includegraphics[width=0.32\textwidth]{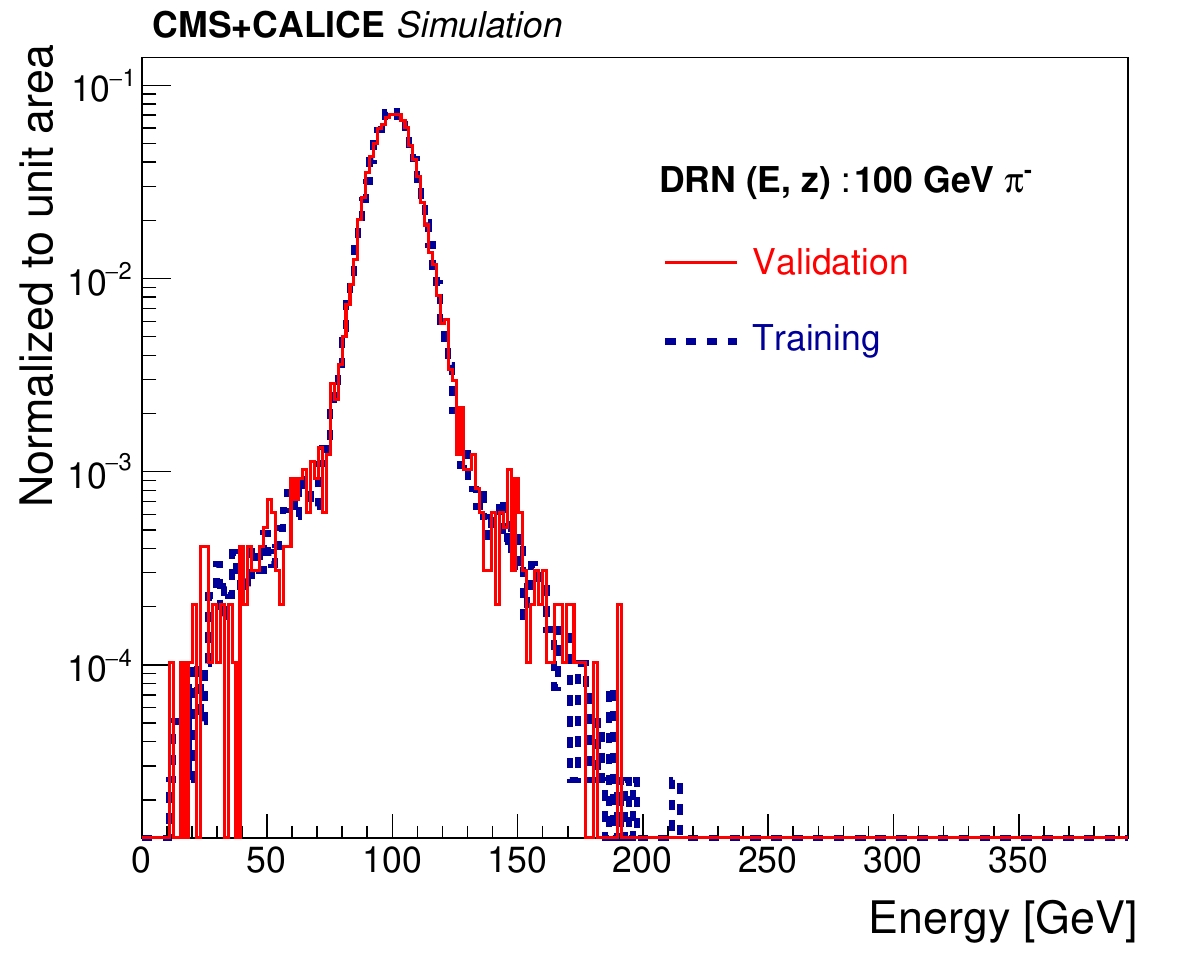}
    \includegraphics[width=0.32\textwidth]{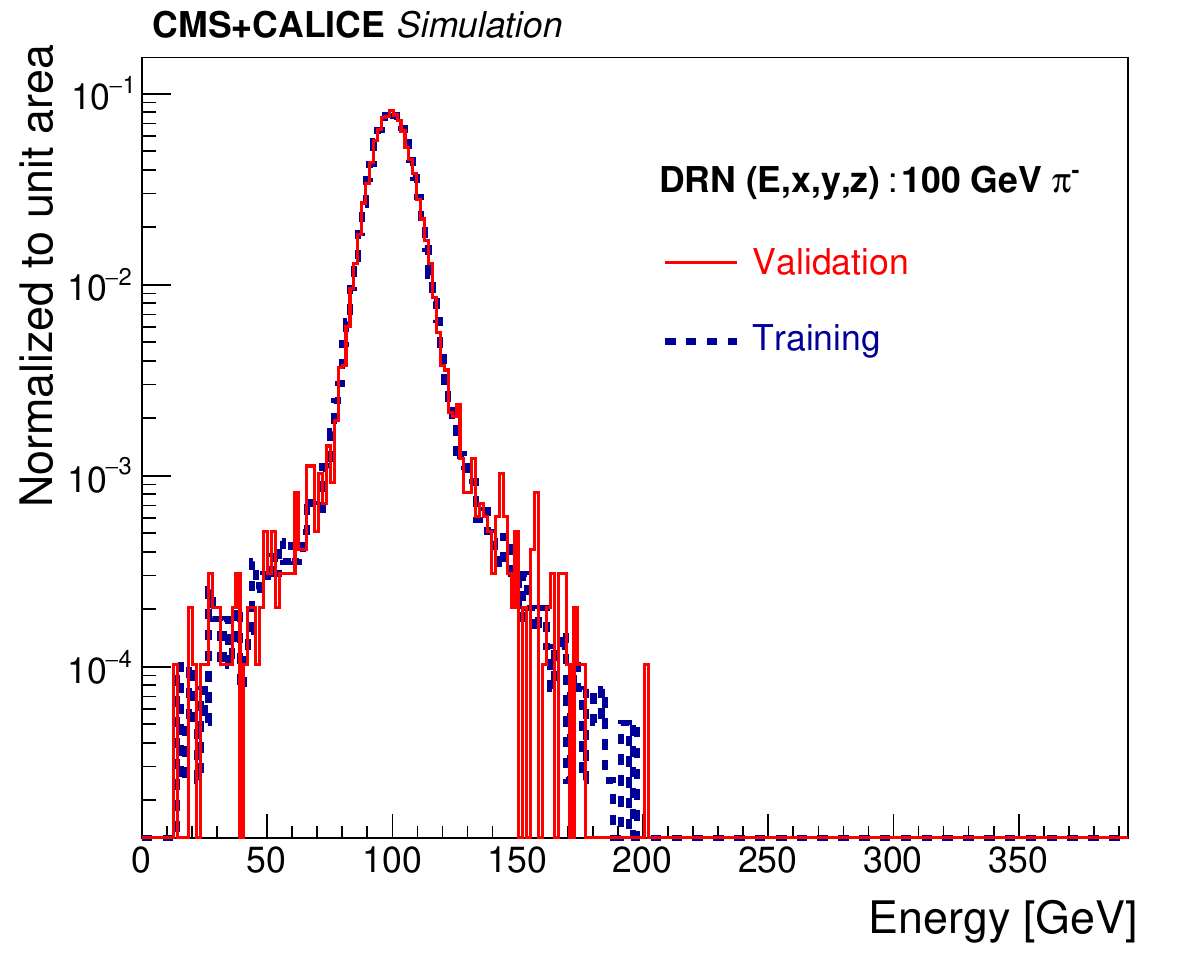}
    \caption{Comparison of distribution of energy obtained in training and validation datasets for the models DRN (E) (left), DRN (E,z) (middle), and DRN (E,x,y,z) (right) for pions with incident energy of 100 GeV.} \label{fig:drn-train-valid-1d}
\end{figure}

\begin{figure}[ht]
    \centering
    \includegraphics[width=0.30\textwidth]{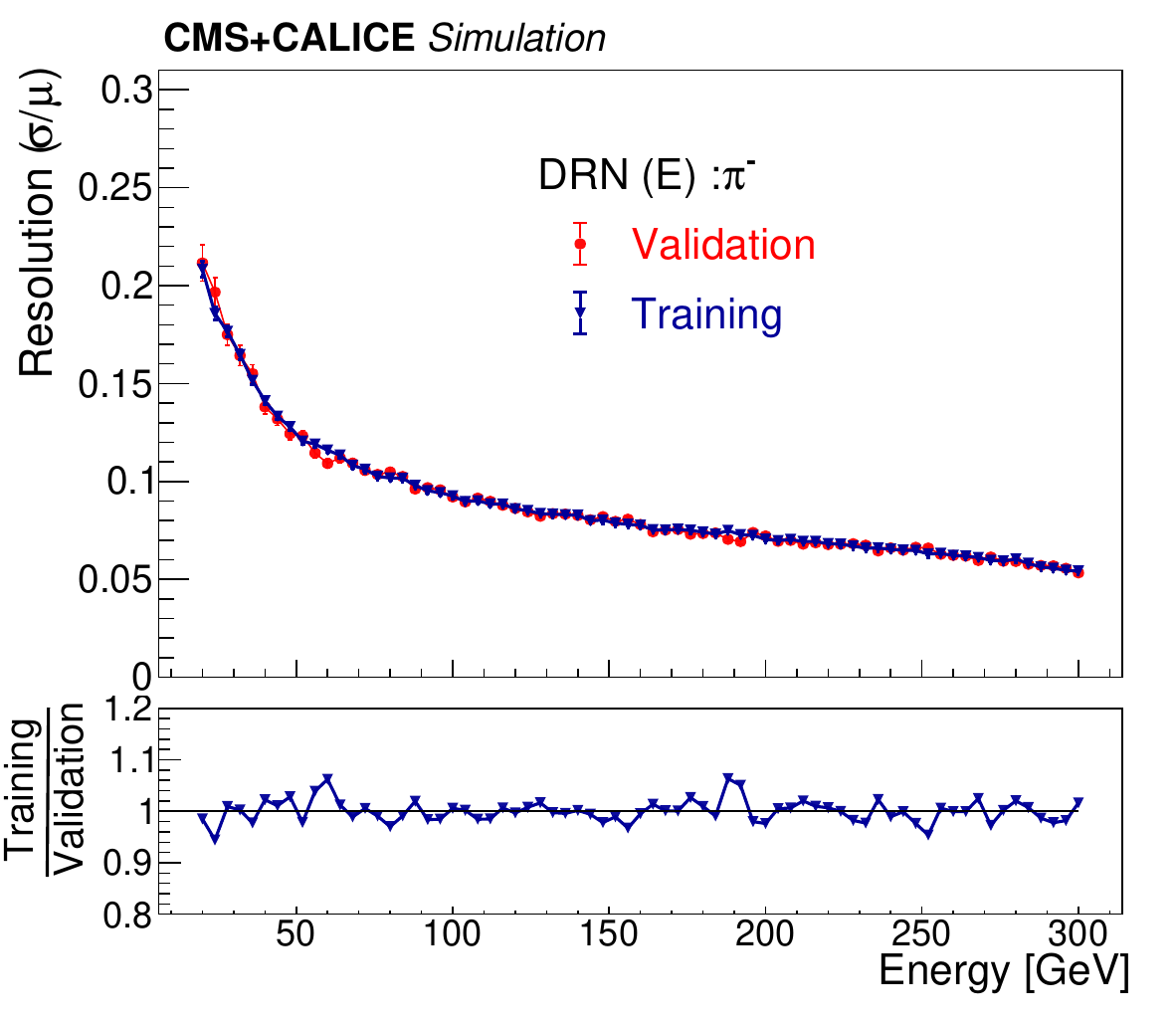}
    \includegraphics[width=0.30\textwidth]{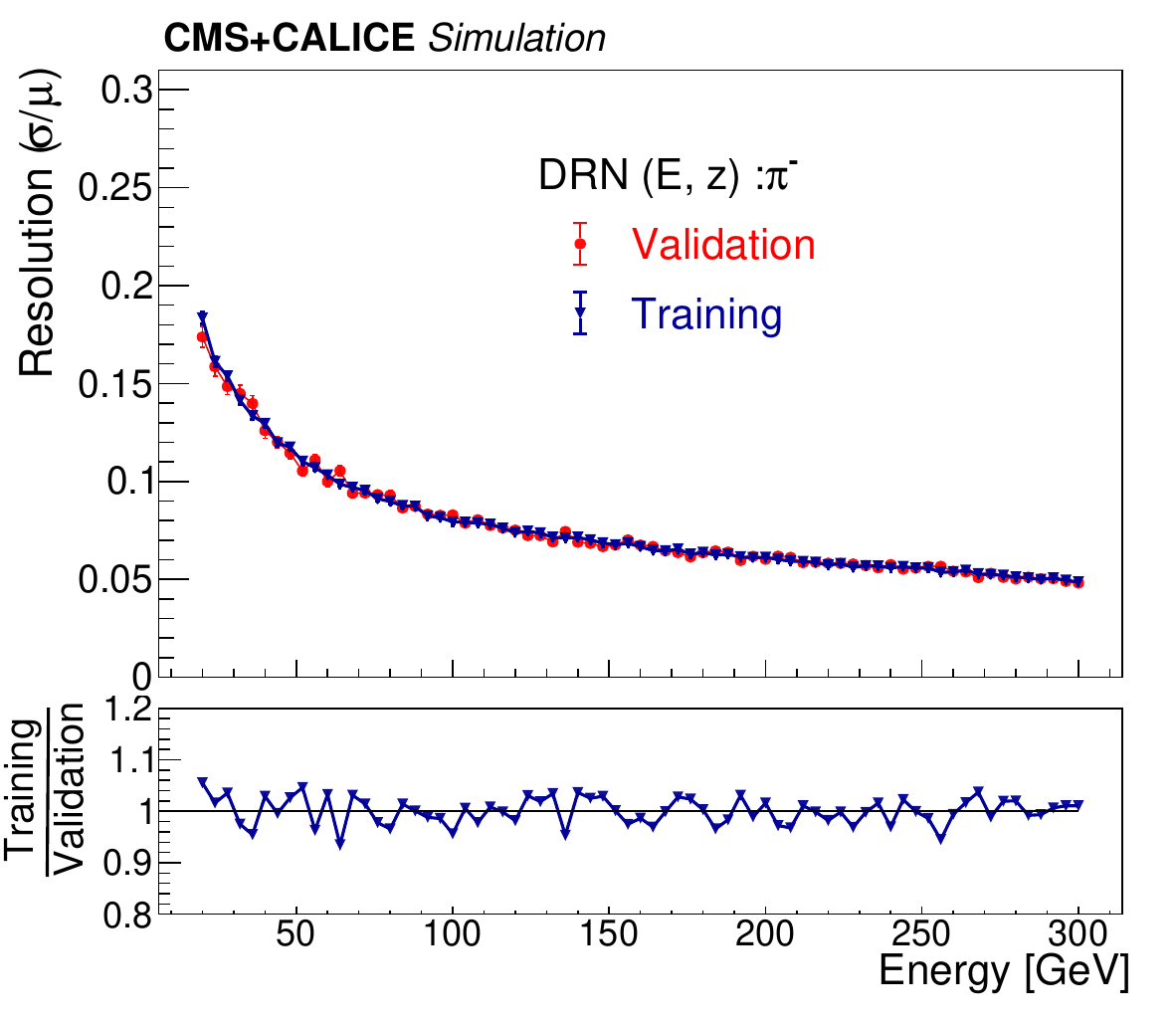}
    \includegraphics[width=0.30\textwidth]{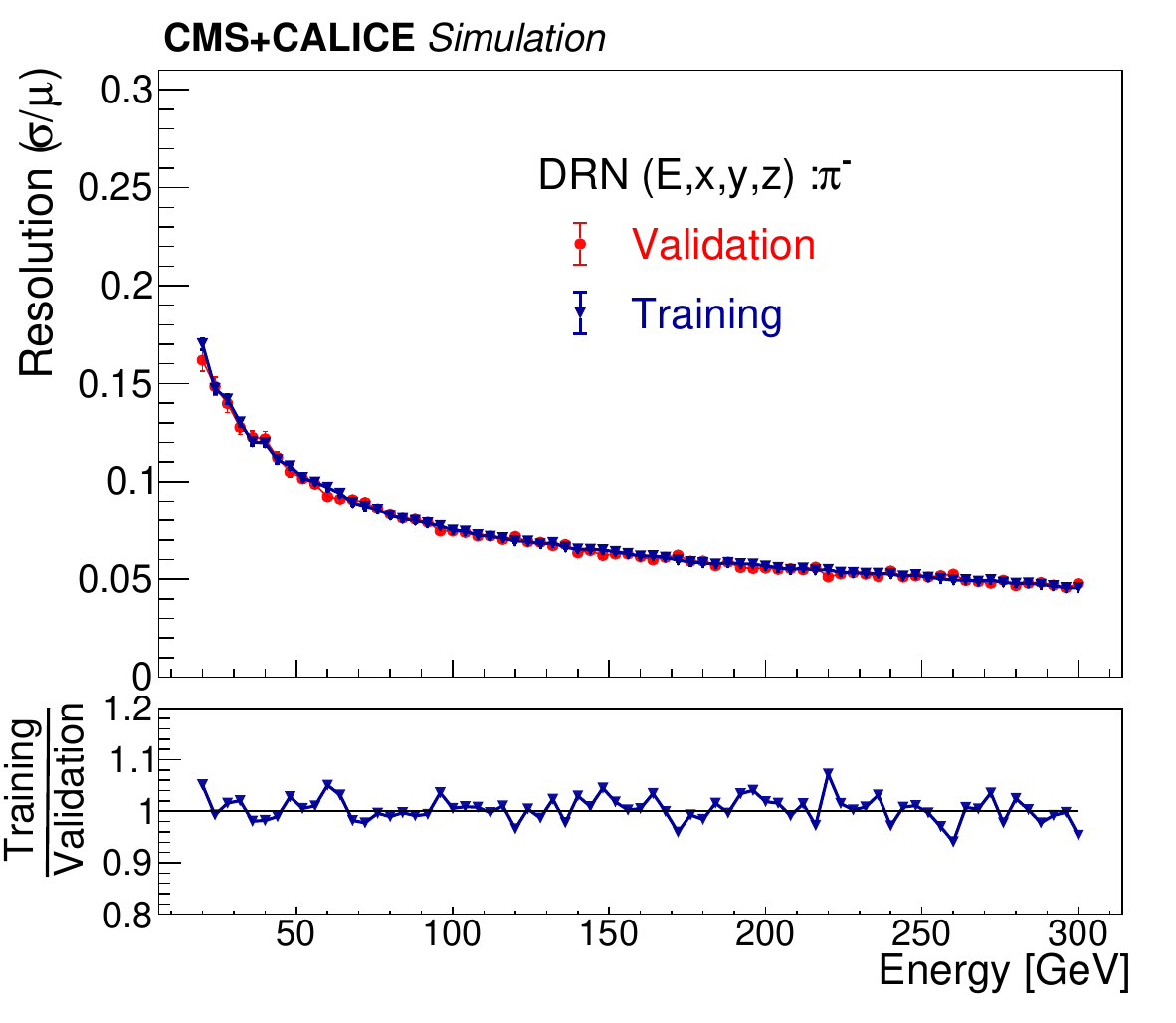} \\
    \includegraphics[width=0.30\textwidth]{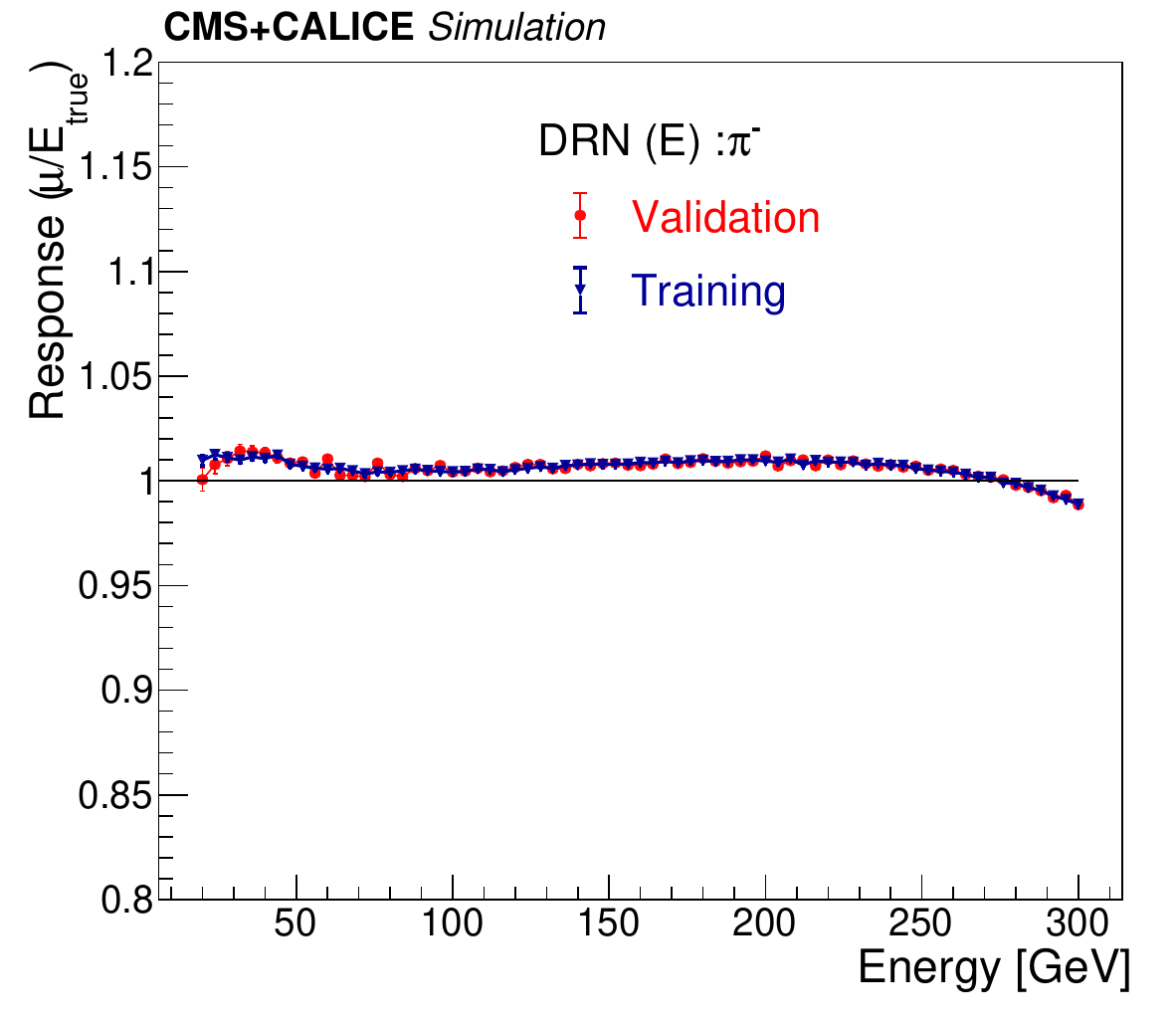}
    \includegraphics[width=0.30\textwidth]{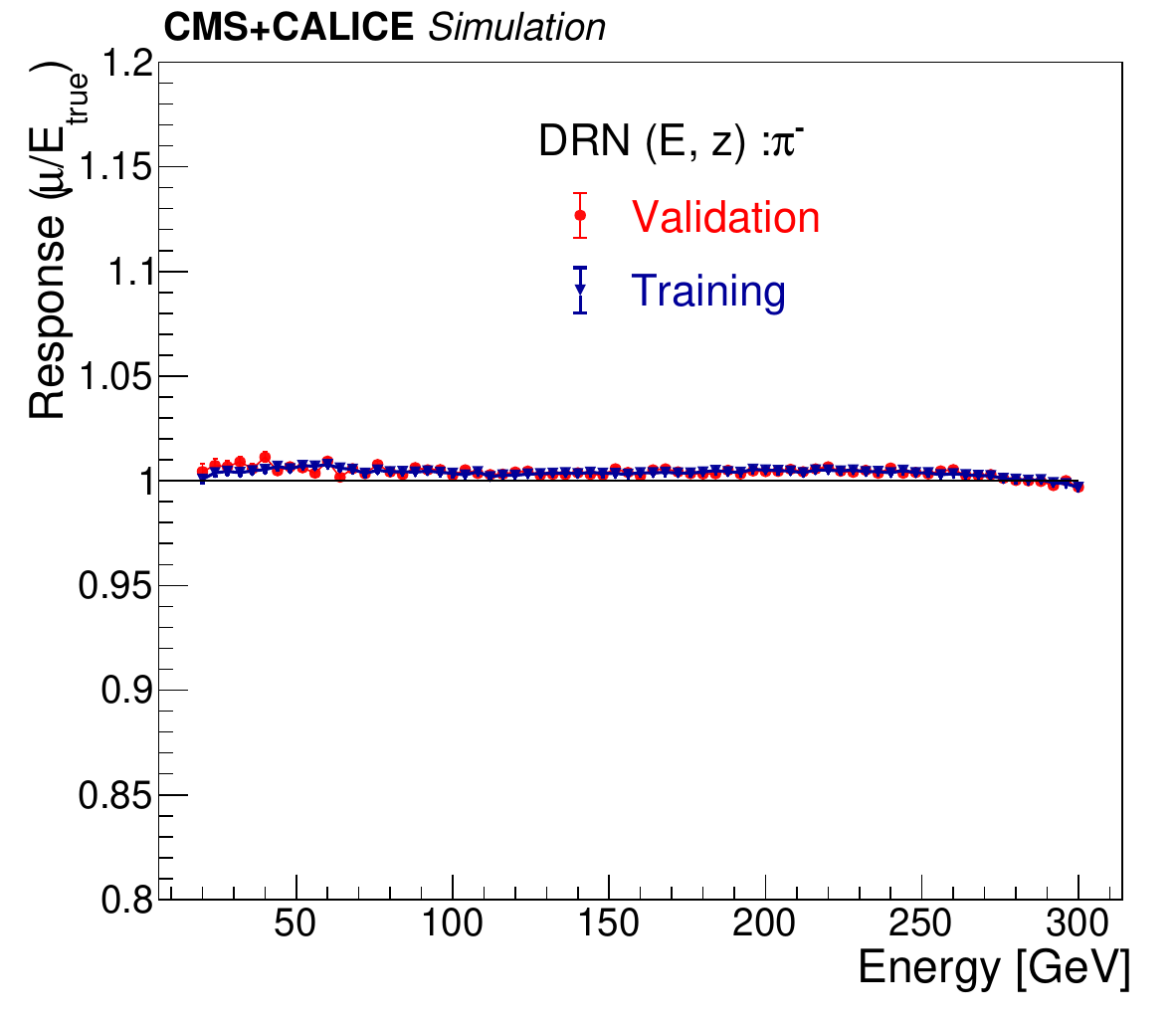}
    \includegraphics[width=0.30\textwidth]{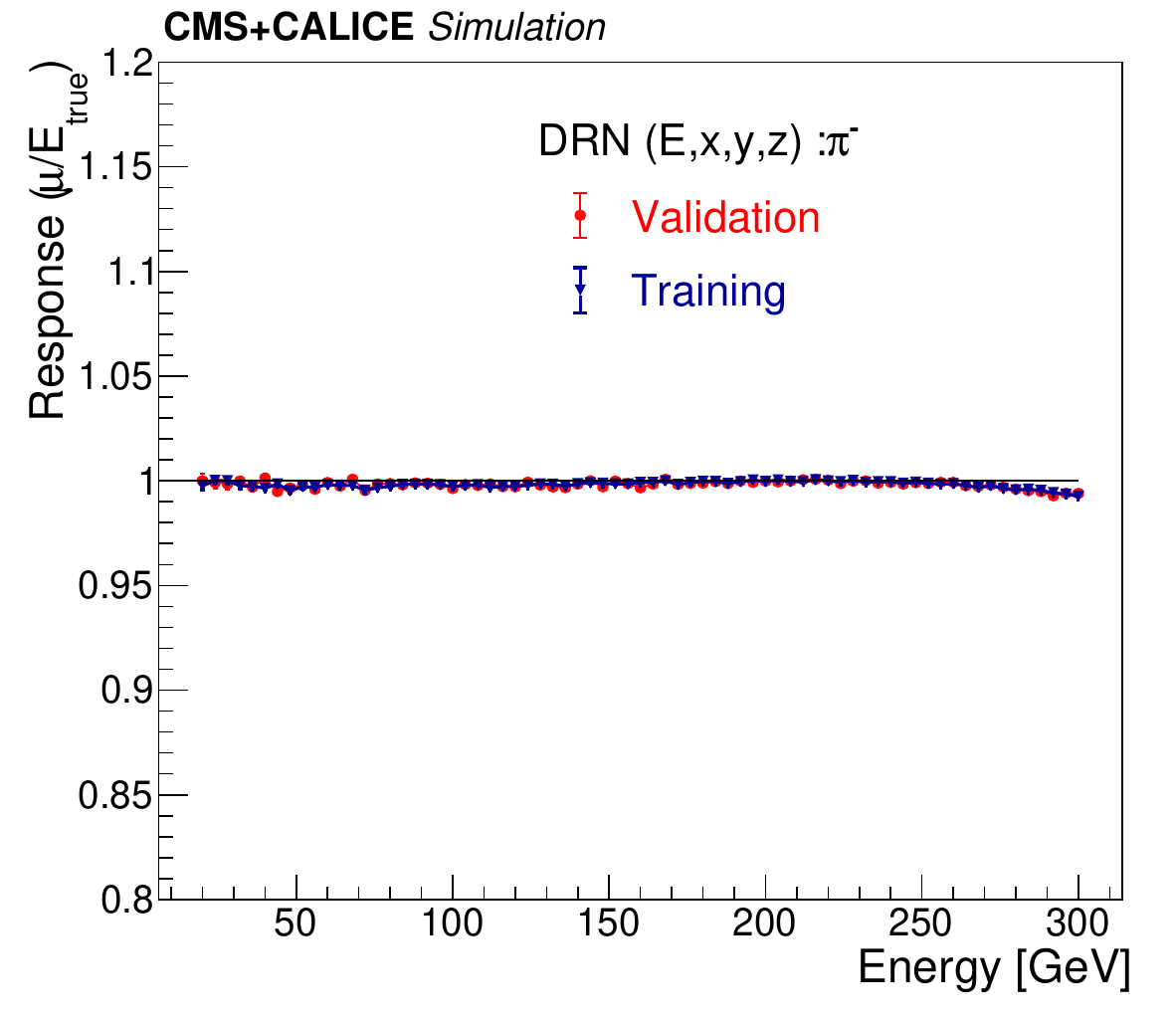}
    \caption{Comparison of resolution (upper row) and response (bottom row) obtained in training and validation datasets for the models DRN (E) (left), DRN (E,z) (middle), and DRN (E,x,y,z) (right) as a function of incident pion energy.}
    \label{fig:drn-train-valid}
\end{figure}

Distributions of the total pion energy predicted in training and validation dataset by the three models are compared in Fig.~\ref{fig:drn-train-valid-1d} for 100 GeV pions. The energy distributions obtained using the validation dataset do not show any bias when compared to those obtained using the event sample used for training the network.
In the bins of the true energy of the charged pions, these distributions are fitted with a Gaussian function in the range of one standard deviation around the mean measured energy. 
The energy resolution is then obtained as the ratio of the standard deviation ($\sigma$) to the mean ($\mu$) obtained from the fitted Gaussian function. 
The energy response of pion showers is defined as the $\mu$, normalized to the true energy.
The response and resolution as a function of pion incident energy found with the validation dataset match well with those obtained in the training dataset, as shown in Fig.~\ref{fig:drn-train-valid} for the DRN(E) (left), DRN(E,z) (middle), and DRN(E,x,y,z) (right).
For the remaining studies, we present the results obtained using the hadron showers in the validation event sample unless stated otherwise.

\begin{figure}[ht]
    \centering
    \includegraphics[width=0.45\textwidth]{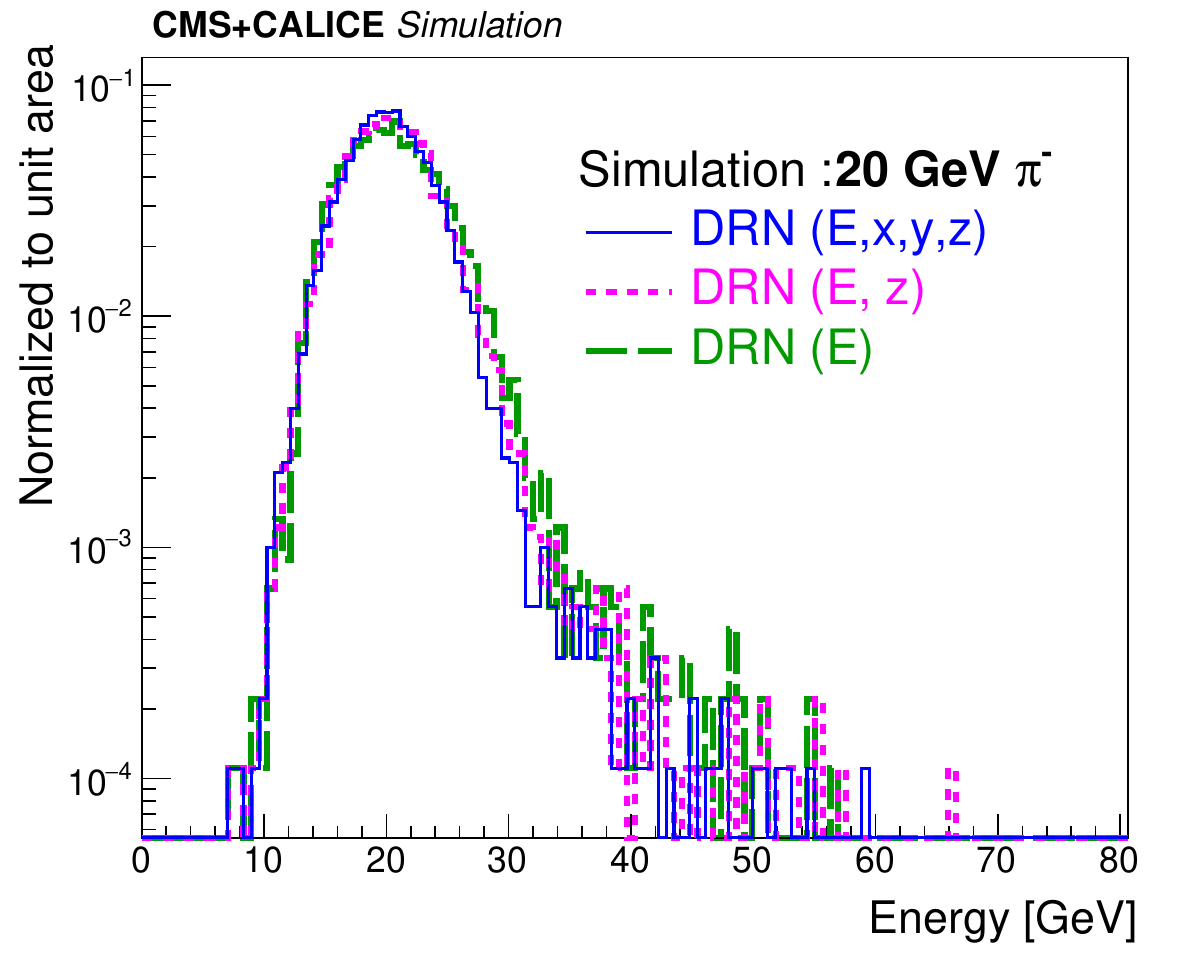}
    \includegraphics[width=0.45\textwidth]{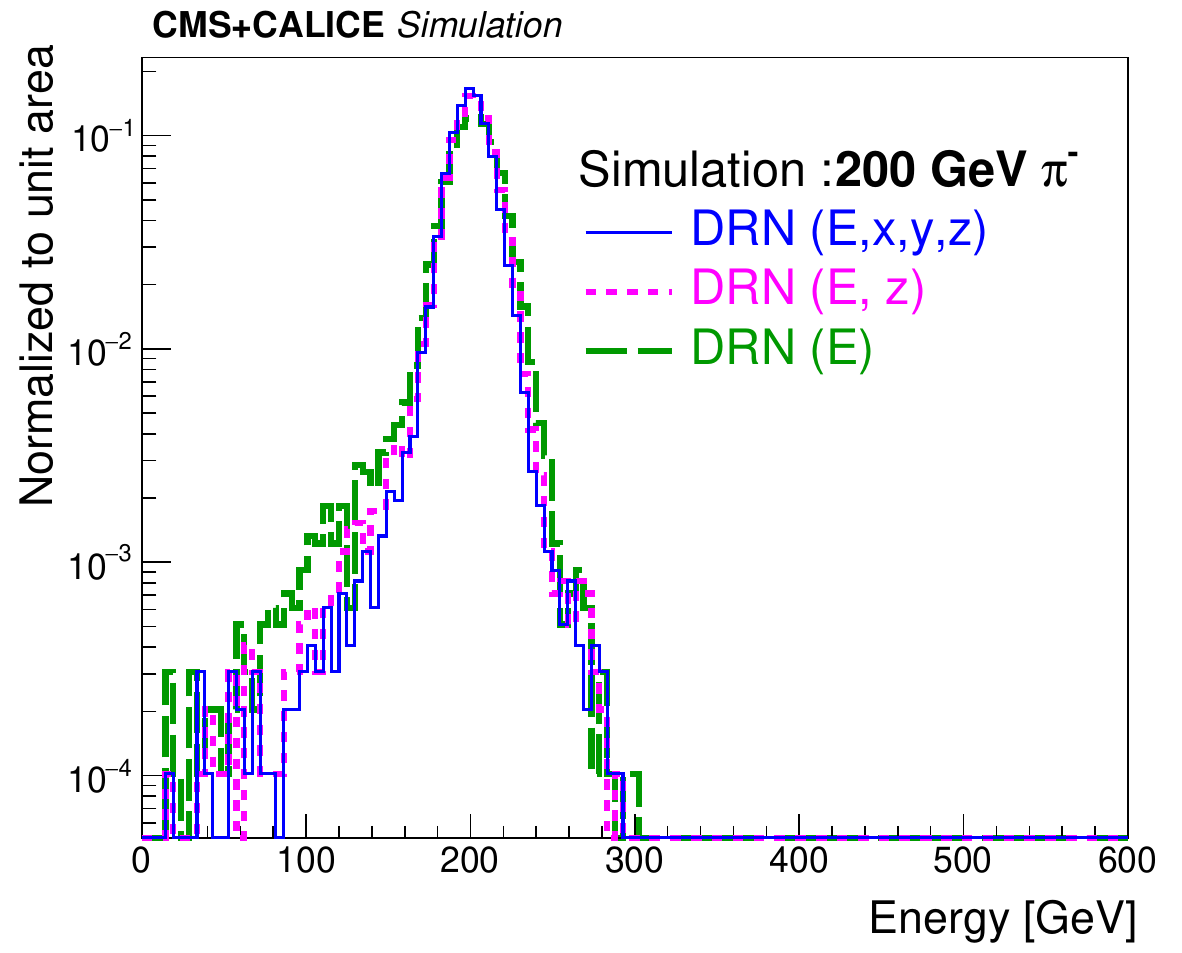}
    \caption{Distributions of energies predicted using different DRN models for 20 GeV (left) and 200 GeV (right) pions.}
    \label{fig:drn-threemodels}
\end{figure}

\begin{figure}[ht]
    \centering
    \includegraphics[width=0.45\textwidth]{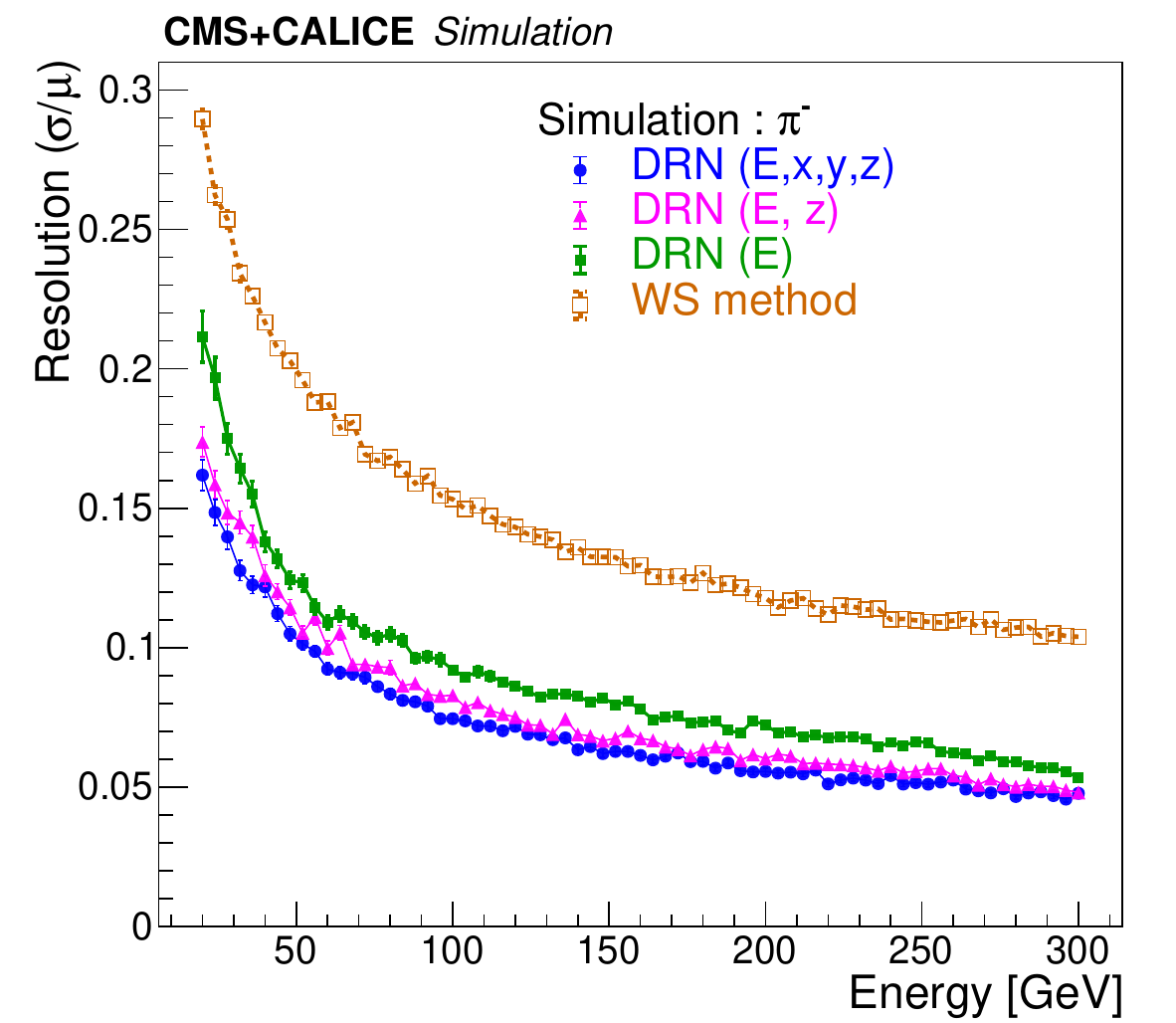}
    \includegraphics[width=0.45\textwidth]{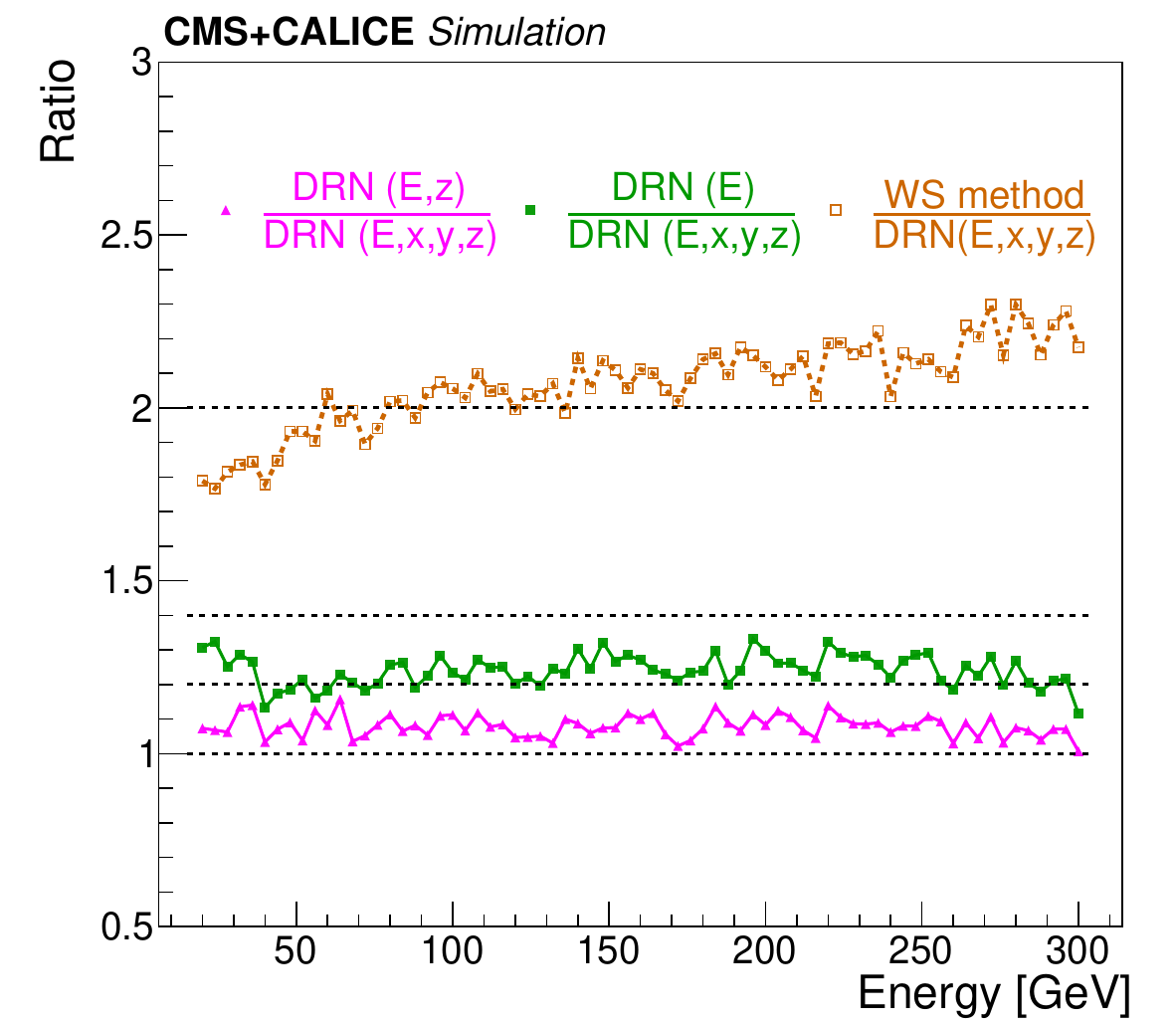}
    \includegraphics[width=0.45\textwidth]{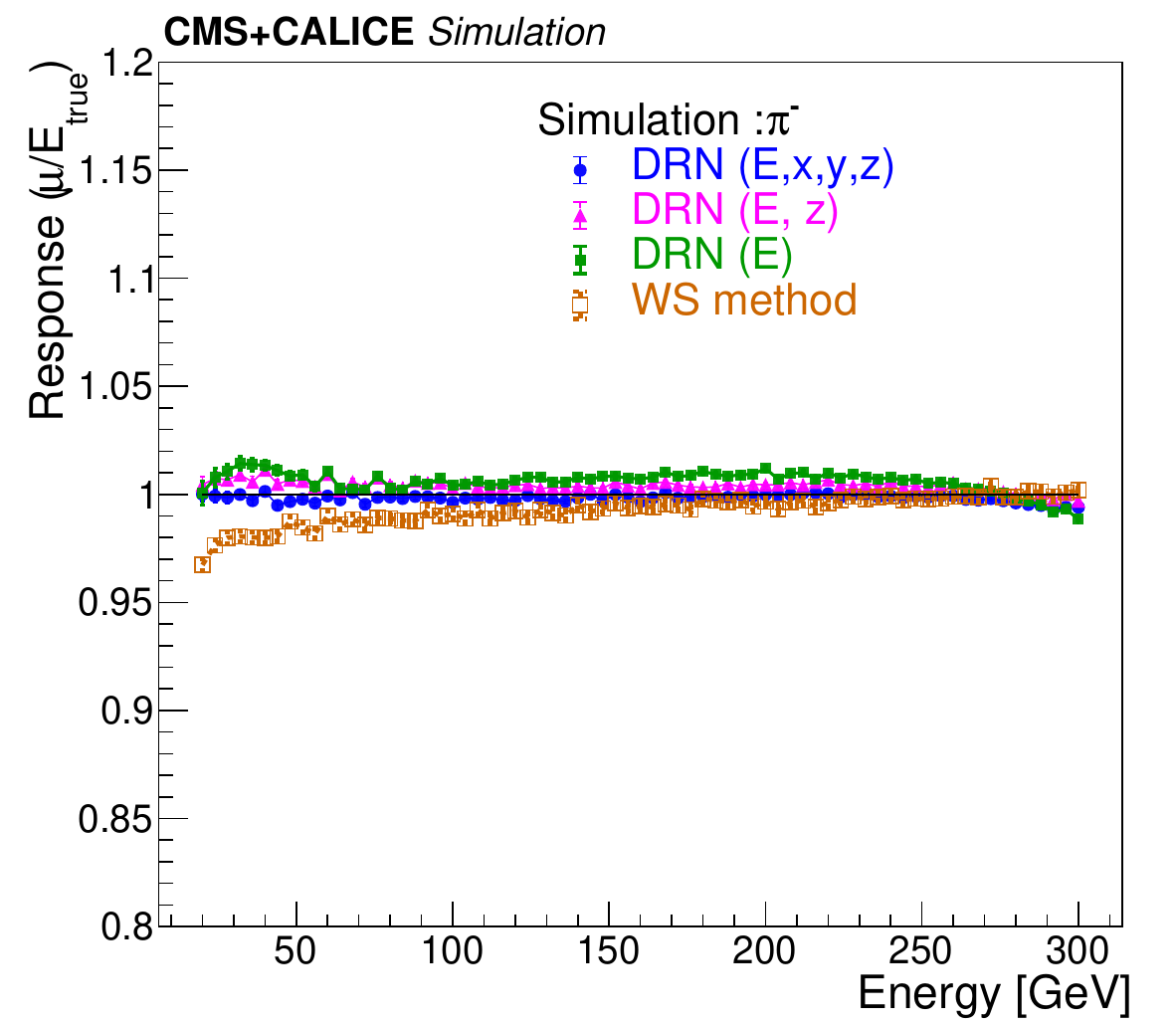}
    \caption{Resolution (top, left), ratio of resolution to DRN (E,x,y,z) (top, right), and response (bottom) as function of pion energy obtained from the three models trained with different features and the WS method in simulation.}
    \label{fig:drn-reso-resp}
\end{figure}

Distributions of energy reconstructed by the three models are shown in Fig.~\ref{fig:drn-threemodels} for pions with different energies. 
We observe that the results obtained restricting to energy information alone are improved when the longitudinal and transverse positions of the rechits are included as input features to the model.  
The resolution as a function of the pion energy obtained from the three models and for comparison the WS method are shown in Fig.~\ref{fig:drn-reso-resp} (top, left). Figure~\ref{fig:drn-reso-resp} (top, right) shows the ratio of resolution obtained using DRN(E), DRN(E,z), and WS method to DRN(E,x,y,z).
There is more than a factor of two improvement in the resolution of energy reconstructed using DRN (E) and an additional 20\% improvement when the longitudinal information of rechits is added.
Adding transverse position information further improves the resolution by up to 10\%. 
These sequential improvements can be interpreted as an indication that the network has learned how the shower spreads in the calorimeter, and is able to partially correct for the energy leakage of the detector.
The energy response obtained using these models is close to the true energy of pions as shown in Fig.~\ref{fig:drn-reso-resp} (bottom).

\begin{figure}[ht]
    \centering
    \includegraphics[width=0.45\textwidth]{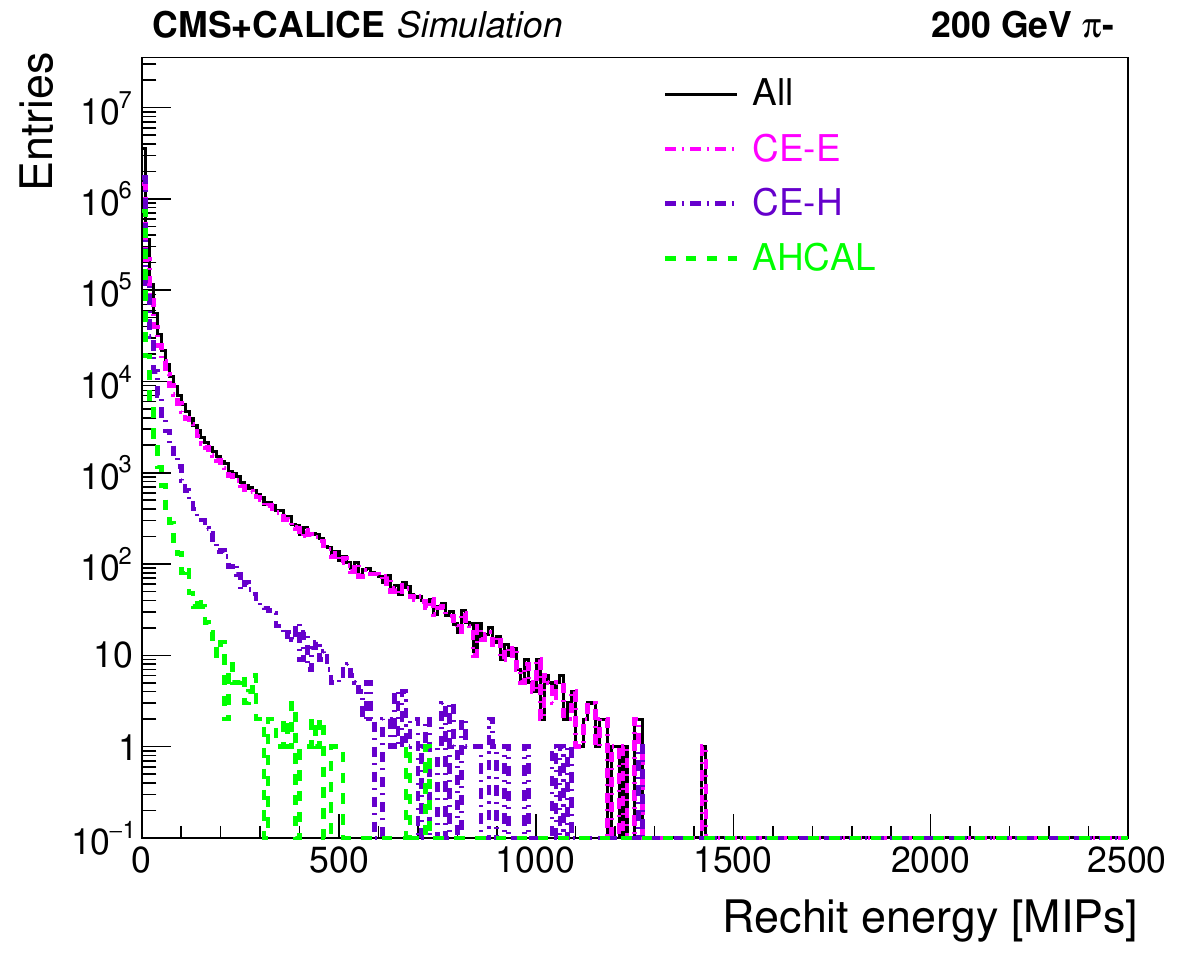}
    \includegraphics[width=0.45\textwidth]{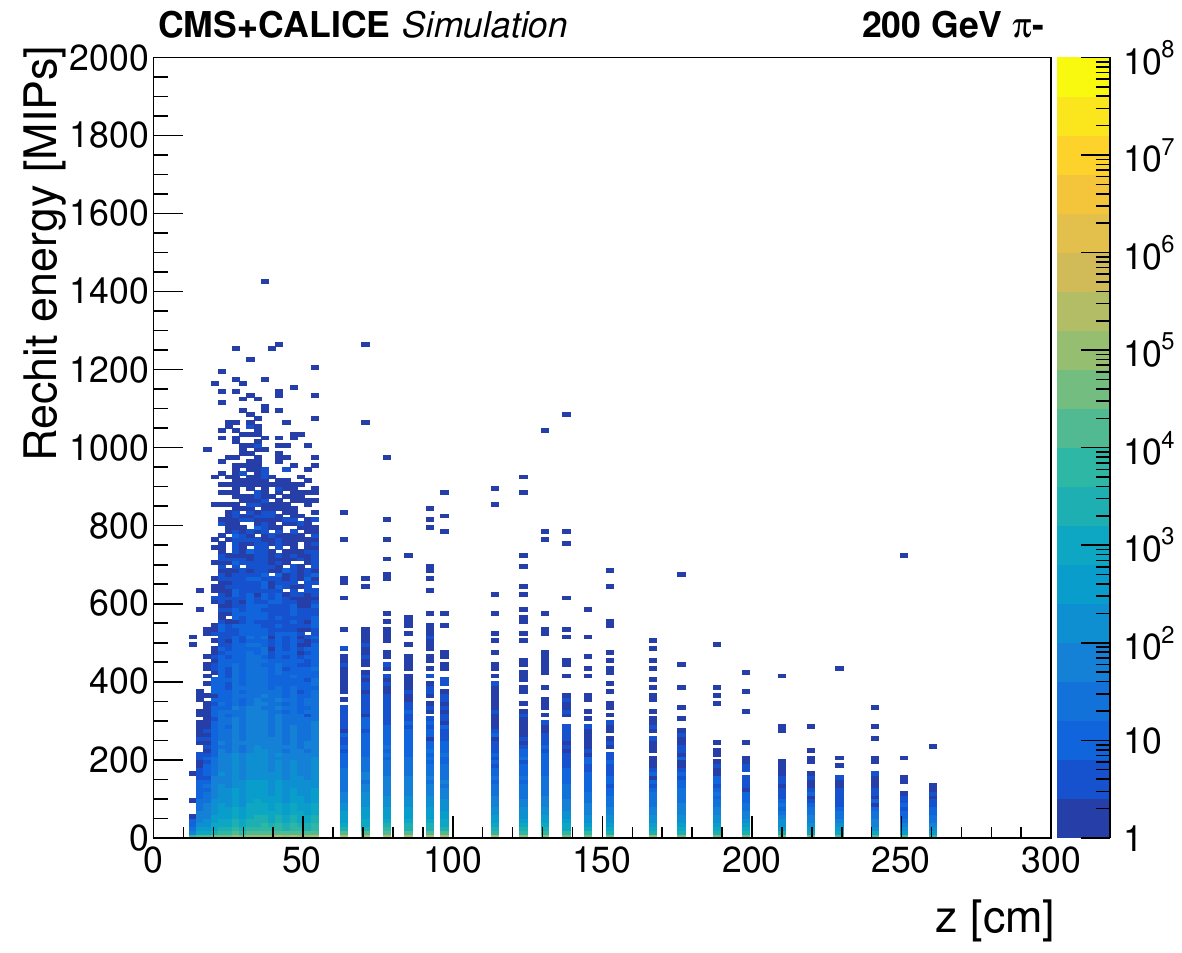}
    \caption{Distributions of rechit energies in units of MIPs (left) and rechit energies as function of depth in the detector (right).}
    \label{fig:drn-input-feat-mips}
\end{figure}

The large improvement in resolution with DRN (E) over the WS method is attributed to the fact that the information of the spatial pattern of energy deposited is encoded in the distribution of rechit energies obtained using detector level calibration.
Distributions of rechit energies measured in units of MIPs in CE-E, CE-H, and AHCAL sections, and their distribution in depth in the detector ($z$ coordinates) are shown in Fig.~\ref{fig:drn-input-feat-mips}.
Comparing the corresponding distributions shown in Fig.~\ref{fig:input-features}, the rechit energies in units of GeV have clusters with much lower energy in CE-E than in the other two sections. 
This is also visible from the much larger MIP-to-GeV conversion for electromagnetic section.

\begin{figure}[ht]
    \centering
    \includegraphics[width=0.65\textwidth]{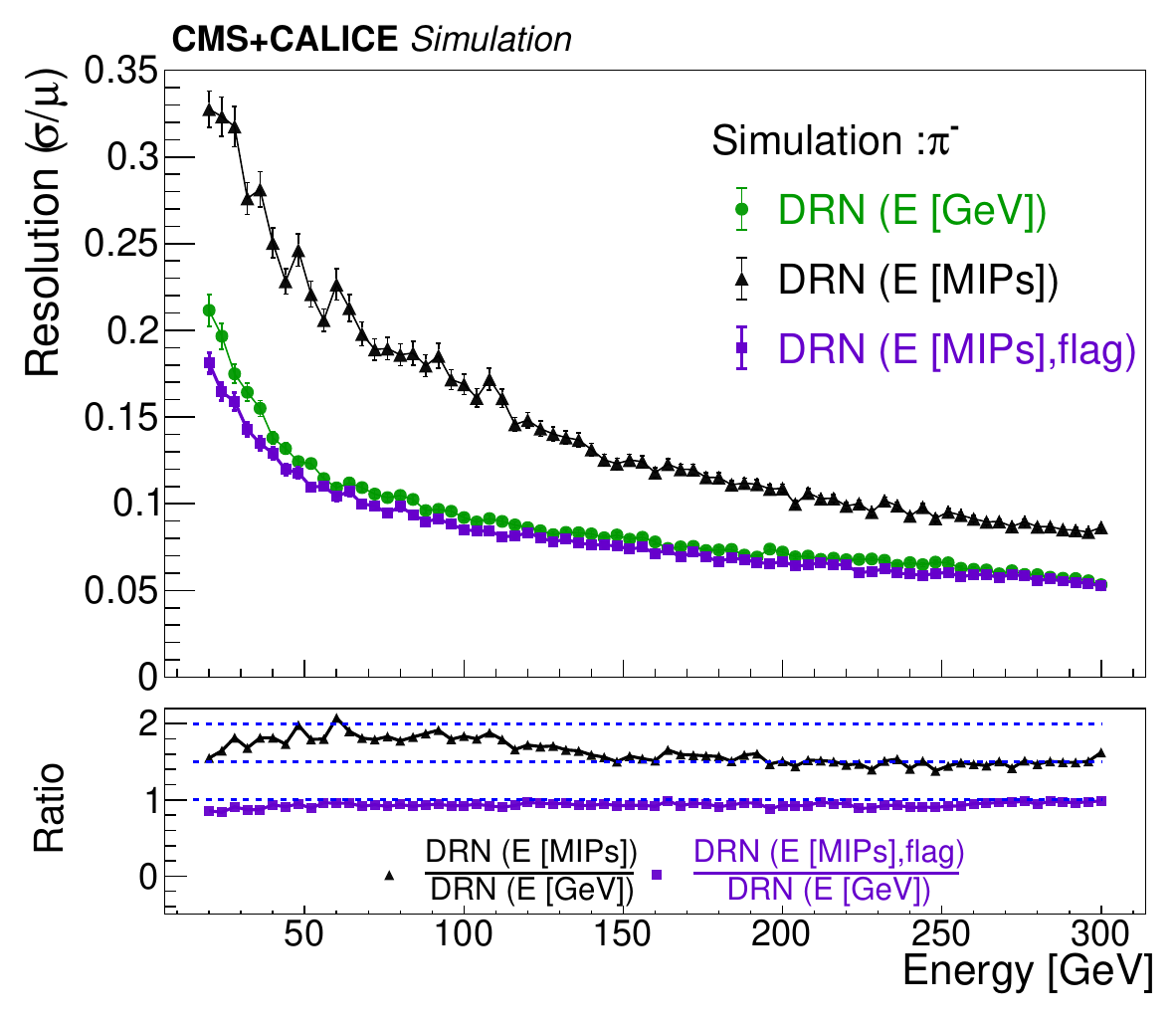}
    \caption{Resolution curves obtained with the models DRN(E), DRN(E[MIPs]), DRN(E[MIPs],Flag). By including a flag indicating in which subsection of the calorimeter the rechit belongs, the resolution obtained with DRN(E) is recovered.}
    \label{fig:drn-mips-flag-respreso}
\end{figure}

To study the capability of the DRN to correct for relative miscalibration of different sections of the calorimeter with different absorber materials and very different thicknesses we trained a DRN(E[MIPs]) model where the rechit signal magnitudes were expressed in units of MIPs instead of GeV. 
We also trained a second DRN(E[MIPs],Flag) model with the rechits again in units of MIPs but now with an additional flag to indicate in which section, CE-E, CE-H or AHCAL, the rechit was located, without providing any further longitudinal information.
The resolution curves obtained with these models are shown in Fig.~\ref{fig:drn-mips-flag-respreso}.
The DRN (E[MIPs]) model performs much worse than DRN (E), whereas the performance of DRN (E[MIPs],Flag) model is similar to, or slightly better than DRN (E) model. 

\begin{figure}[ht]
    \centering
    \includegraphics[width=0.45\textwidth]{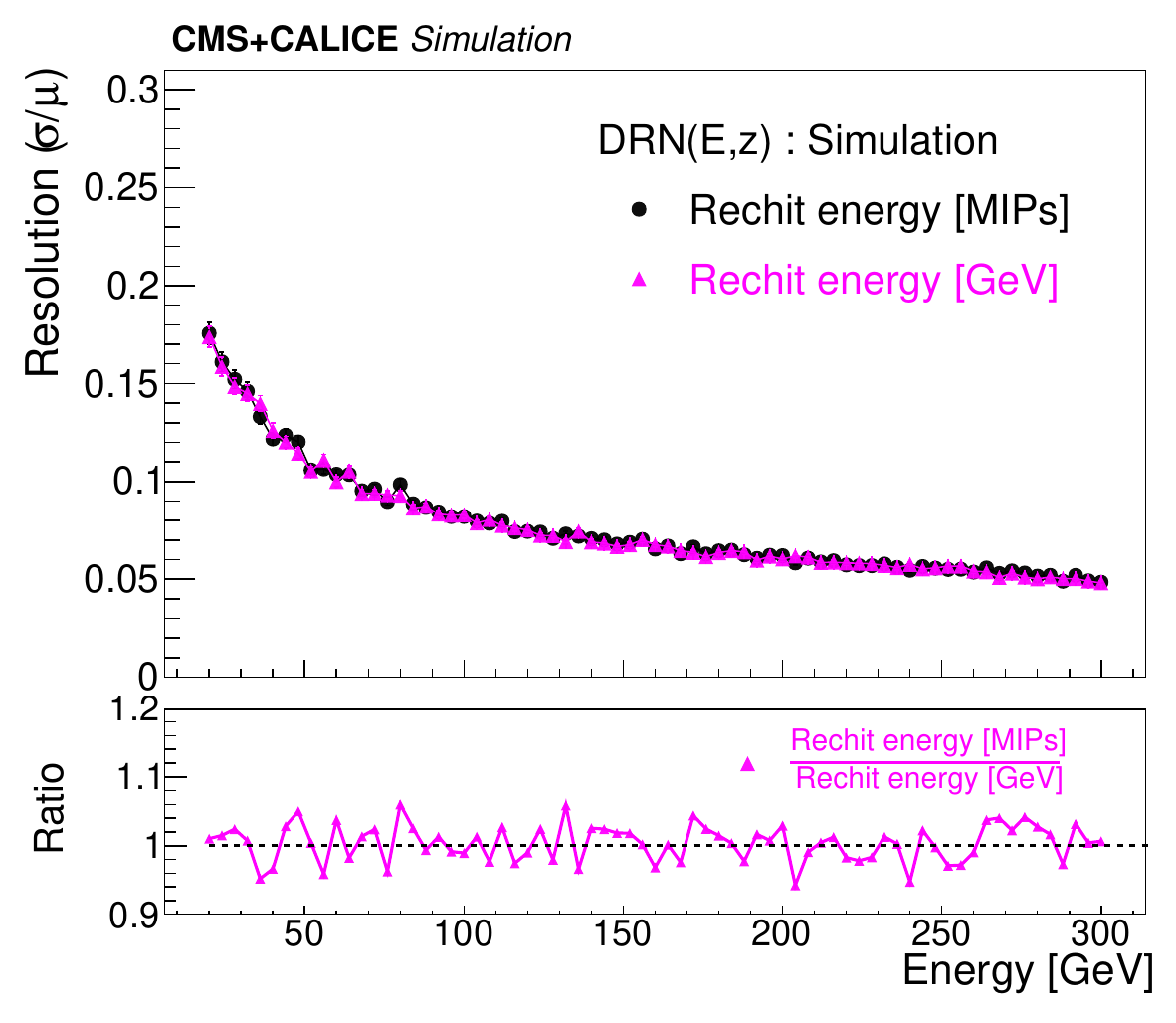}
    \includegraphics[width=0.45\textwidth]{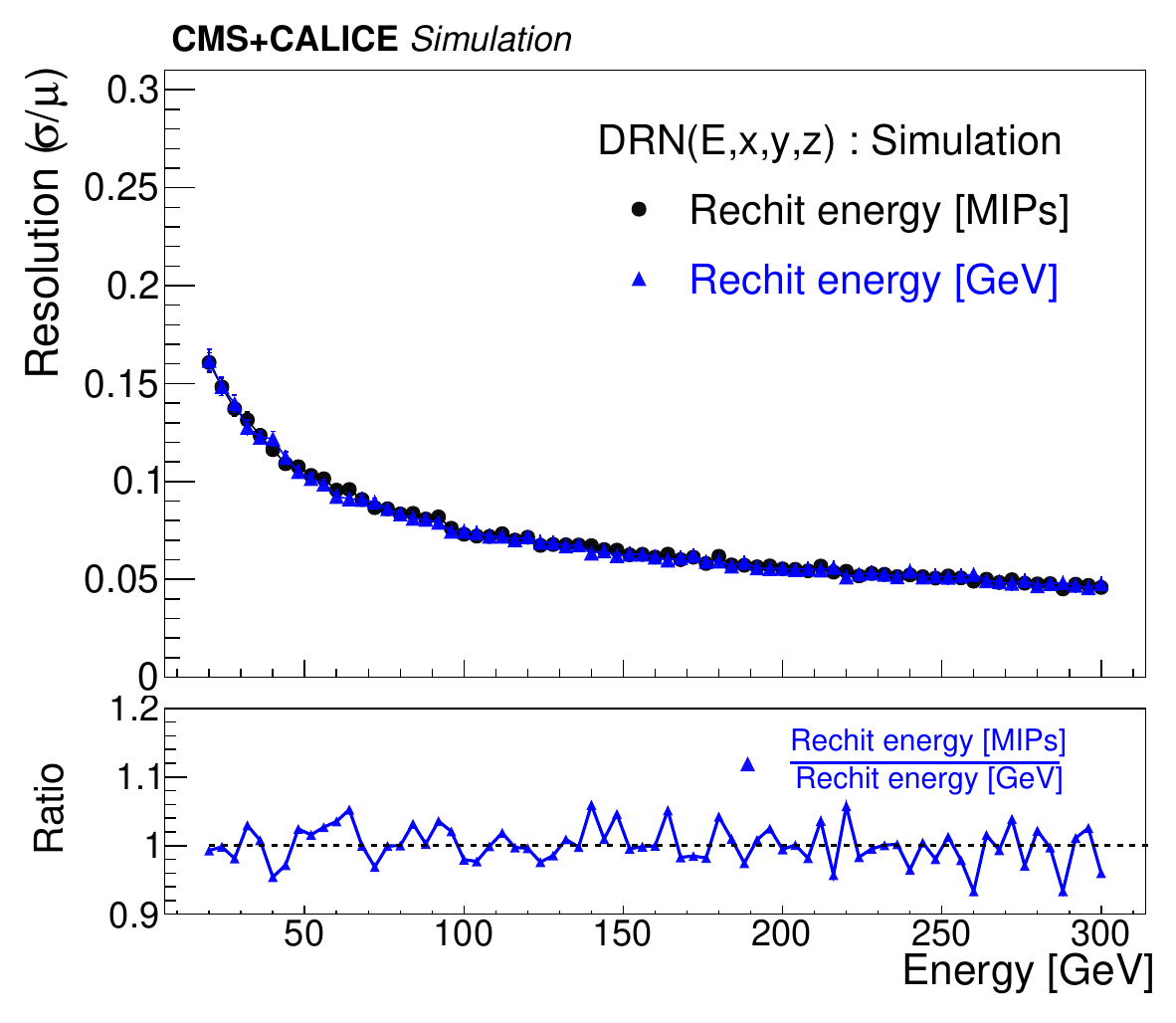}
    \caption{Comparison of resolution using rechit energies given in GeV or MIPs with only longitudinal (left) and both transverse and longitudinal (right) coordinates.}
    \label{fig:drn-mips-flags-ezxy}
\end{figure}

Figure~\ref{fig:drn-mips-flags-ezxy} shows the energy resolution using E[MIPs] along with the $z$ and $x$,$y$,$z$ information. 
We find that providing the spatial information of rechits makes the details of rechit energy calibration redundant and the performance of the DRNs trained on different sets of input features is similar. 

\section{Performance of DRN with beam test data}\label{sec:drn-tbdata-sim}

We tested the DRNs with real data collected with a negatively charged pion beam at the CERN SPS in 2018. 
The apparatus is described in Section \ref{sec:tb2018}. 
Event samples of pions were collected with beam energy settings of 20, 50, 80, 100, 120, 200, 250 and 300 GeV. 
The expected purity of the beam was 80--90\% with the balance consisting mostly of muons, based on a detailed simulation of the beam line~\cite{ShubhamThesis}.
The sample size for each energy setting ranged from 80\,k$-$100\,k events collected and 40\,k$-$80\,k events are available after fiducial and other quality selections for further analysis.
The selection criteria are described in Ref.~\cite{bib.hgcal-2018-pions}. 
Approximately 100k events were simulated with \GEANTfour for each of the eight energy points using the same geometry and the same physics lists that were used for the training samples.
The same event selection criteria applied to events in the data were applied to the simulated events.
These fixed energy samples are not used for model training. 

\begin{figure}[ht]
    \centering
    \includegraphics[width=0.40\textwidth]{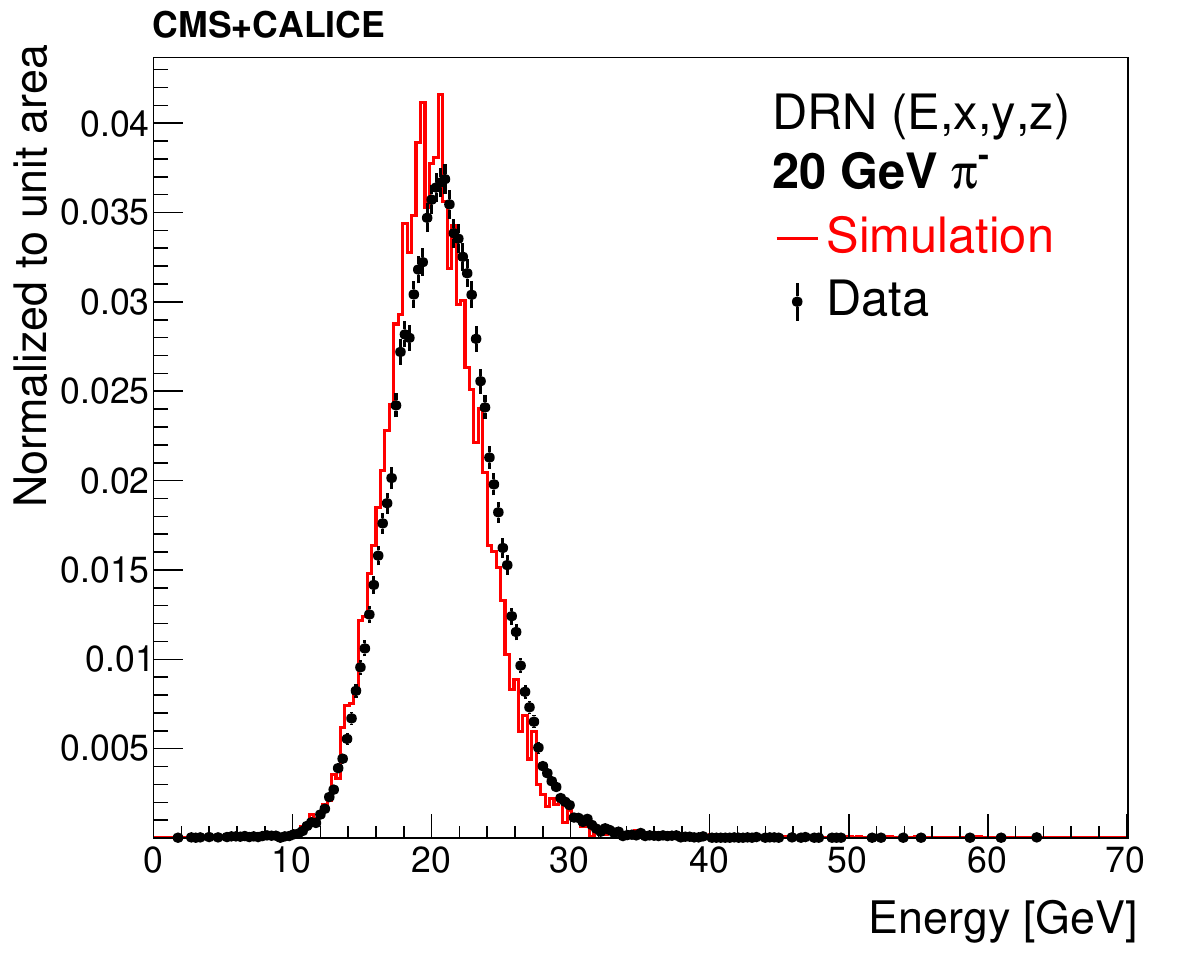}
    \includegraphics[width=0.40\textwidth]{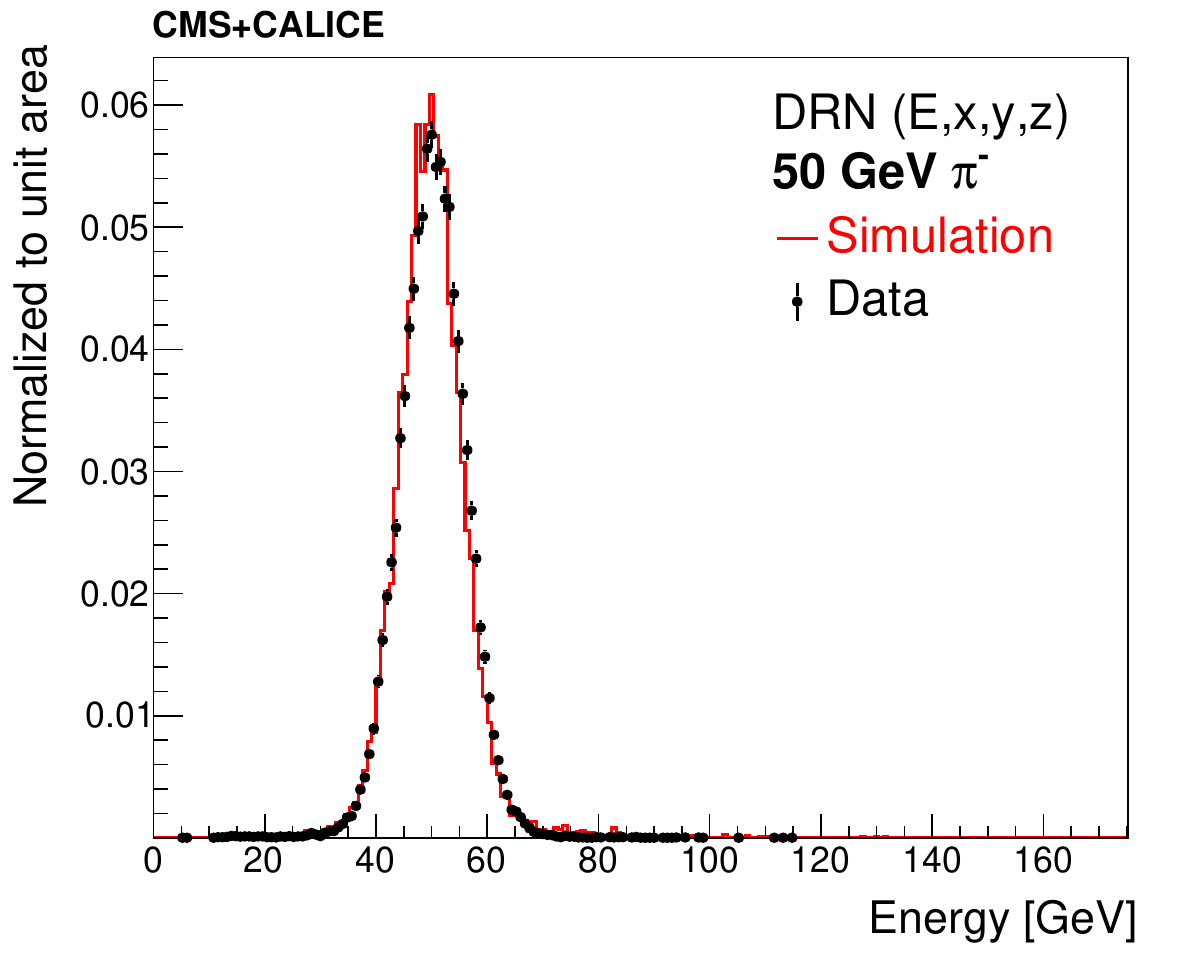}
    \includegraphics[width=0.40\textwidth]{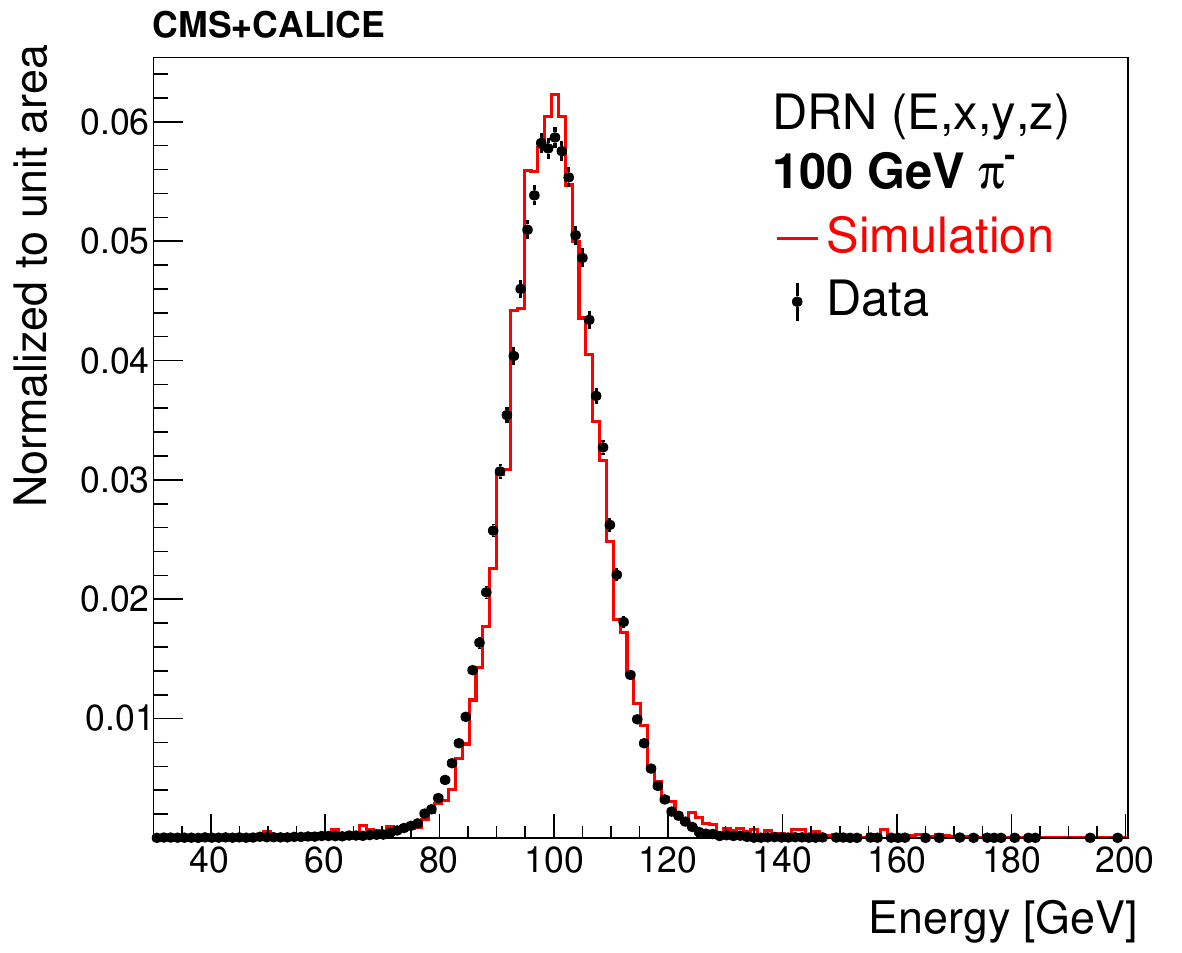}
    \includegraphics[width=0.40\textwidth]{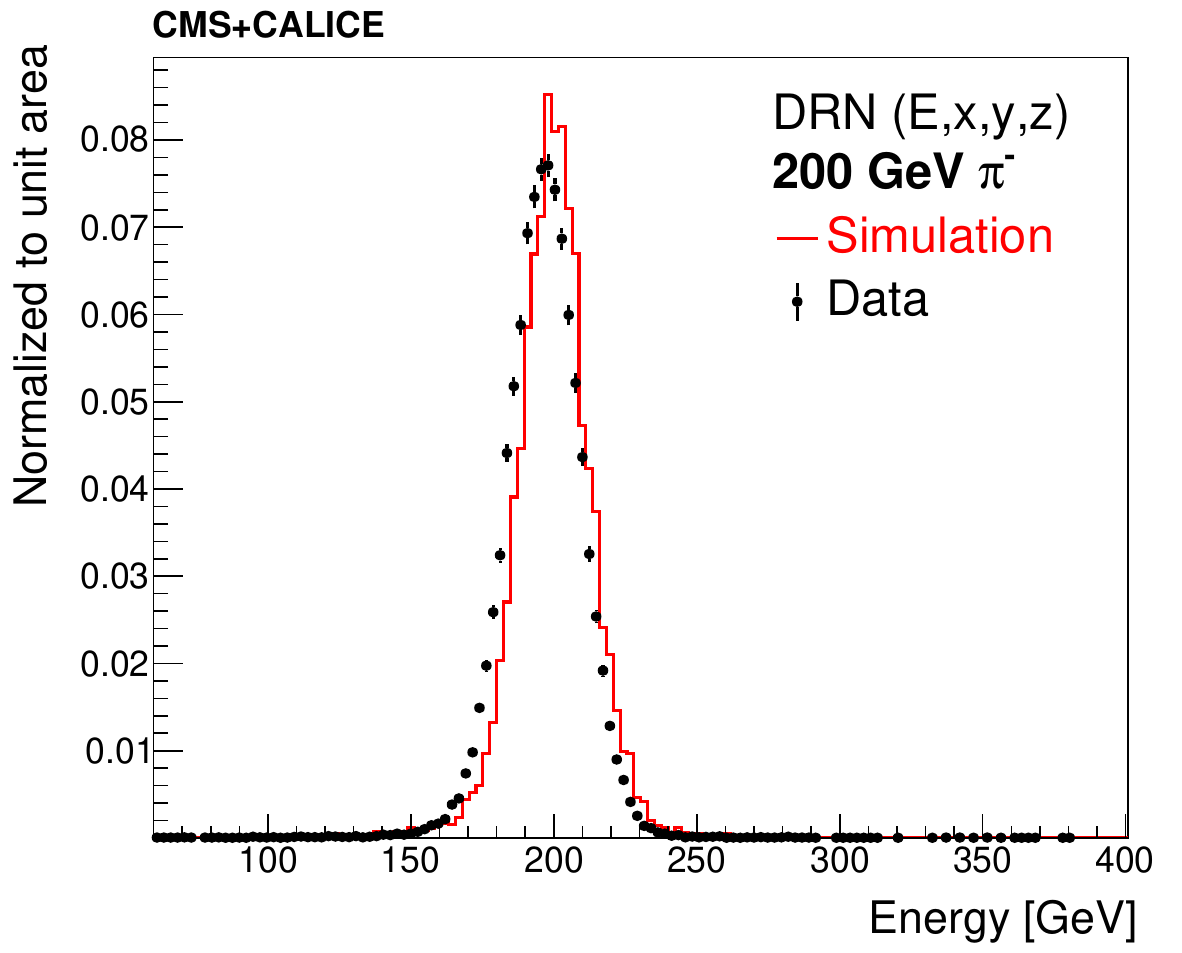}
    \caption{Distribution of energy predicted by the DRN model on hadron showers generated by pions of true energy 20 GeV (top, left), 50 GeV (top, right), 100 GeV (bottom, left), and 200 GeV (bottom, right) in beam test HGCAL prototype detector and its simulation. The model is trained using the simulated flat energy sample in the same detector setup.}
    \label{fig:drn-datasim-1d}
\end{figure}

\begin{figure}[ht]
    \centering
    \includegraphics[width=0.49\textwidth]{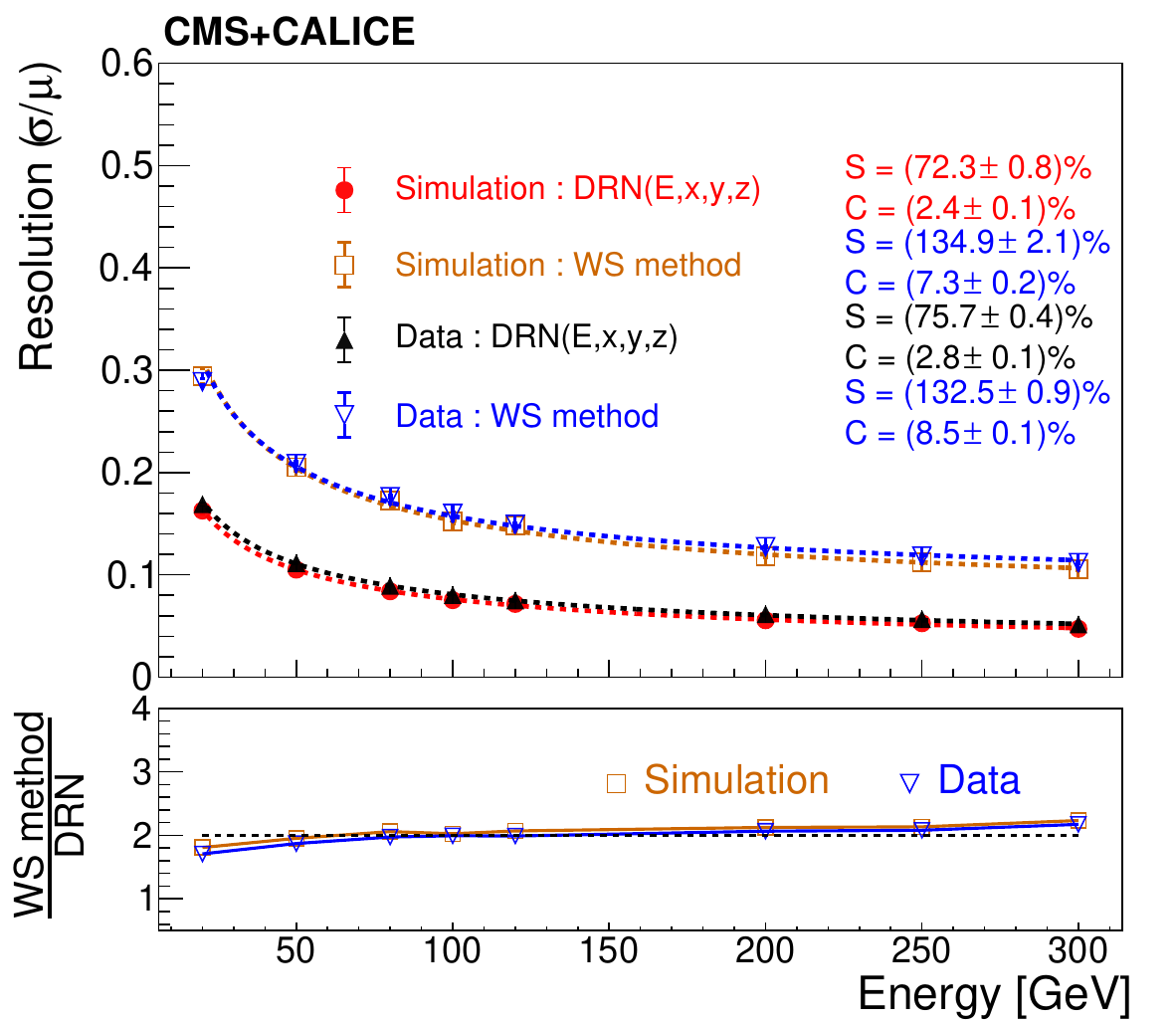}
    \includegraphics[width=0.49\textwidth]{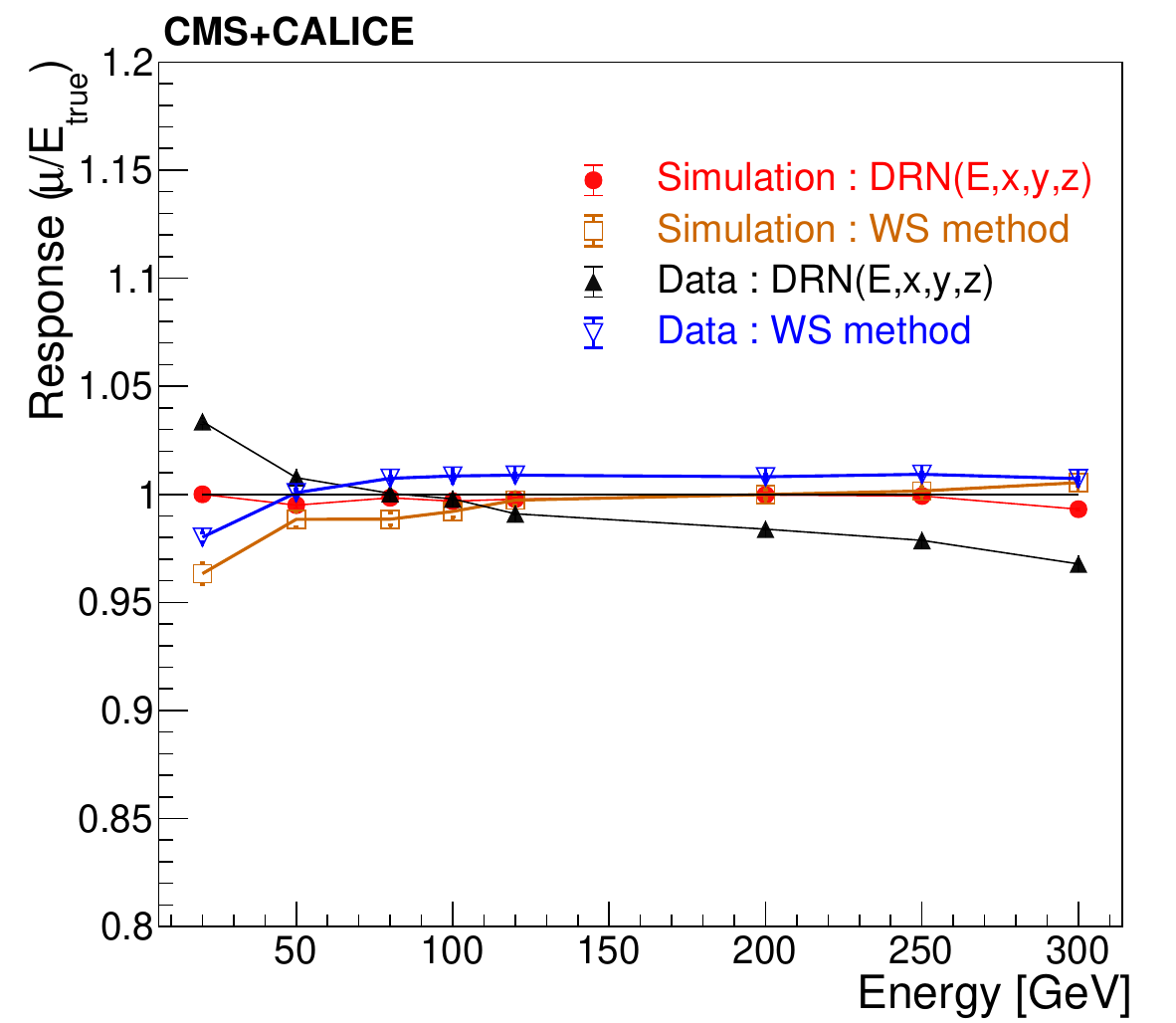}
    \caption{Comparison of resolution (left) and response (right) for pion showers reconstructed in beam test data and simulation using DRN model and WS method.}
    \label{fig:drn-datasim-respreso}
\end{figure}

For comparison of the detector response between beam test data and simulation, we used the DRN (E,x,y,z) model which gave the best results.
We also used the WS method for comparison.
It is identical to the one described in the Ref.~\cite{bib.hgcal-2018-pions} and is re-evaluated for the configuration with a reduced number AHCAL layers that we used.
Comparisons of the energy distribution found using the DRN method in the data and in the simulation are shown in Fig.~\ref{fig:drn-datasim-1d} for beam energies of 20, 50, 100 and 200 GeV, with the model accurately reproducing the overall shape of the measured energy distributions. 
We plot in Fig.~\ref{fig:drn-datasim-respreso} the energy resolution and the energy response as a function of incident energy found with the DRN(E,x,y,z) model and with the WS method.

The DRN model shows similar performance on charged pion showers collected in the beam test experiments and simulated using \GEANTfour.
We use the functional form $\sqrt{S^2/E+C^2}$ to fit the resolution as a function of the incident energy, where $S$ is the stochastic term and $C$ is the constant term. 
We obtain for S and C 75.7\,$\pm$\,0.4\% and 2.8\,$\pm$\,0.1\%, respectively, in data and 72.3\,$\pm$\,0.8\% and 2.4\,$\pm$\,0.1\% in simulation.
We observe an improvement of approximately a factor of two in the resolution to that obtained with the WS method across the whole energy range of the beam test data. 
The energy resolution improves from approximately 20\% (12\%) obtained using the WS method to 10\% (6\%) for 50 (200)GeV pions. 
The response is also within a few-\% of unity. 
The deviation of the response from unity in data is understood to be due to the differences in details of longitudinal and transverse development of shower in the simulation. 
Except for correcting for an overall scale difference~\cite{bib.hgcal-2018-pions}, the simulation has not been tuned to match the fine details in the data. 
Typically in experiments, differences between data and simulation are generally corrected for with scale factors to be determined $\it in~situ$ from the data.

\begin{figure}[ht]
    \centering
    \includegraphics[width=0.49\textwidth]{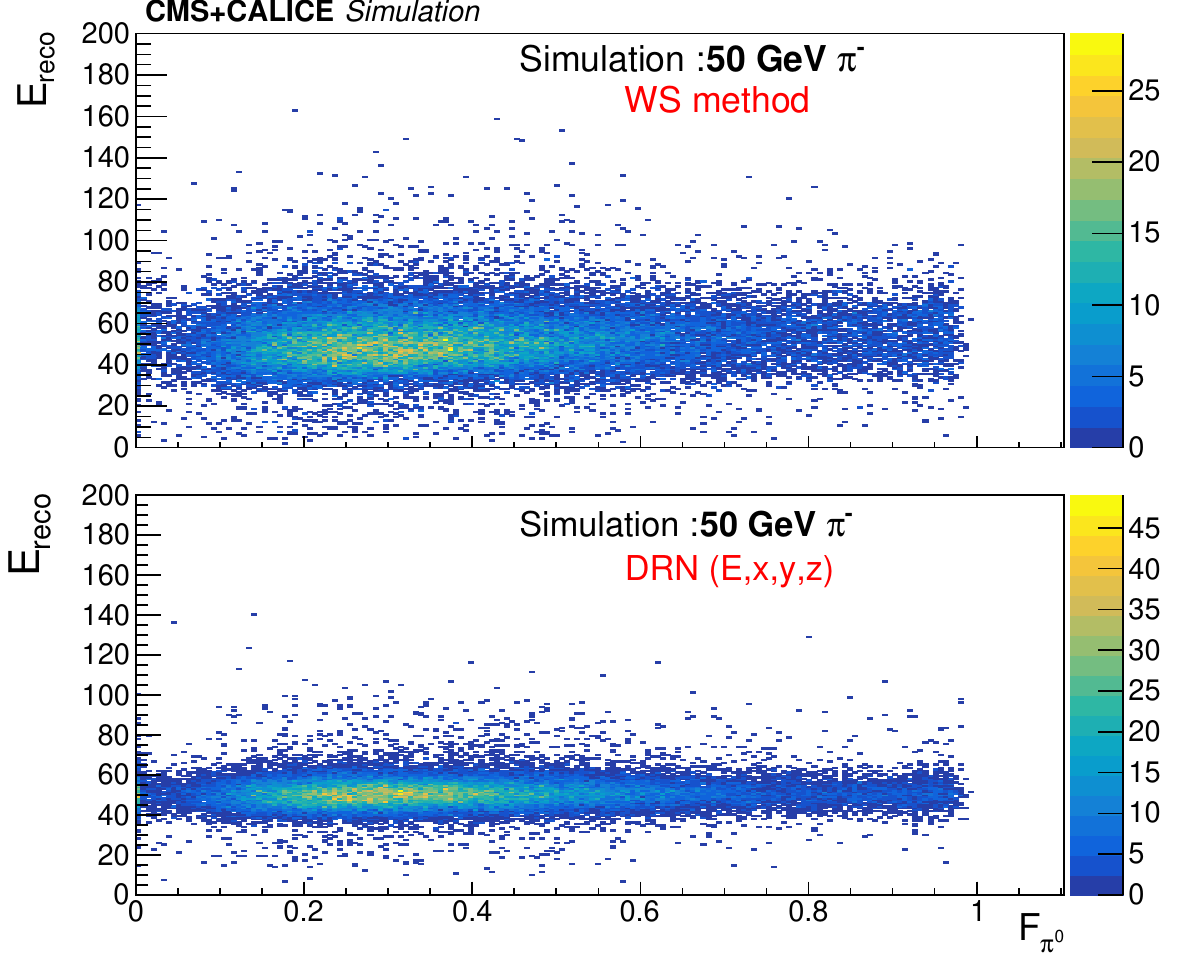}
    \includegraphics[width=0.49\textwidth]{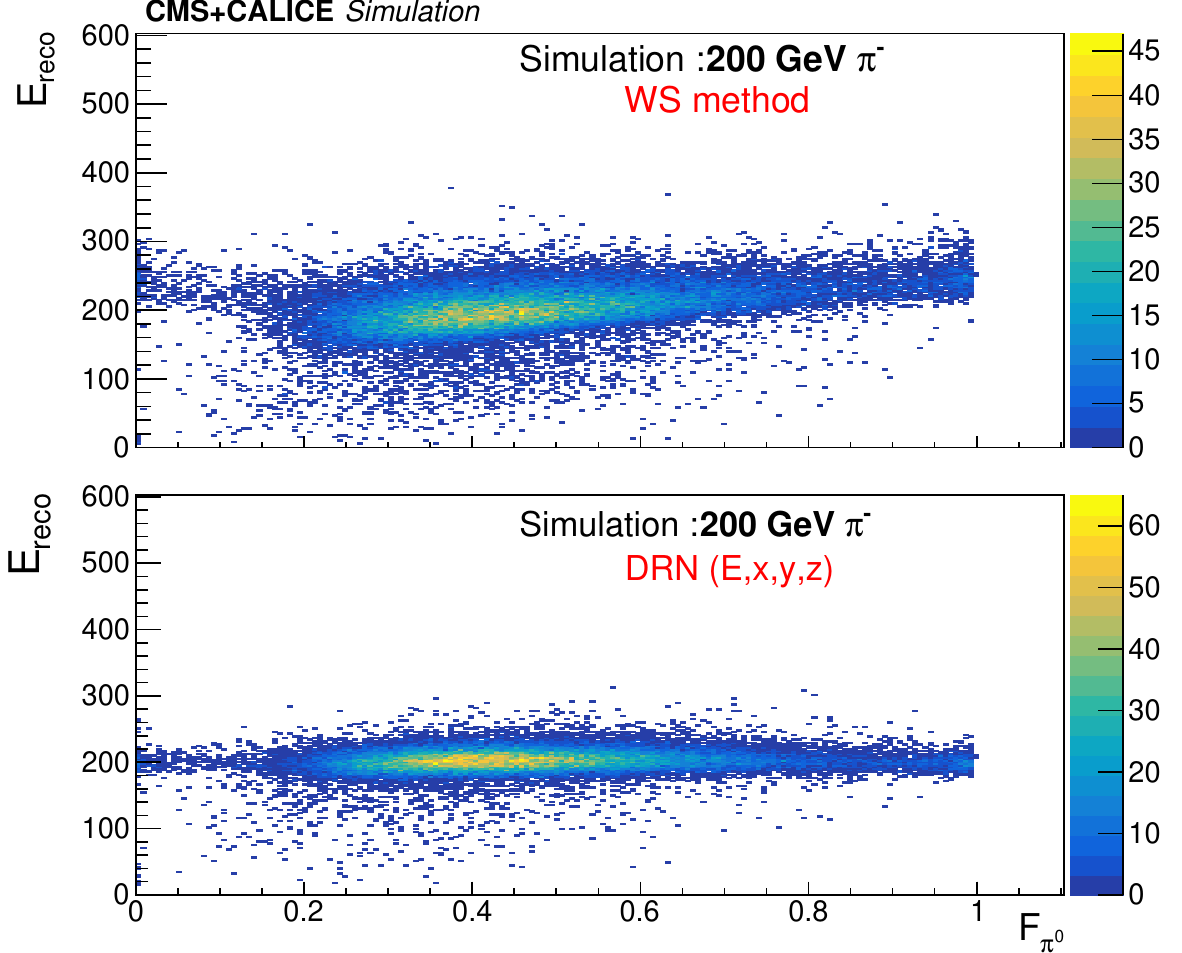}
    \caption{Energy reconstructed using WS method (top panel) and DRN(E,x,y,z) model (lower panel) for pions of energy 50 GeV (left) and 200 GeV (right) as a function of the fraction of energy carried by $\pi^0$s in a given shower.}
    \label{fig:drn-chi2-pi0frac}
\end{figure}

To understand why the DRN significantly improves over the WS method, we note that due to the high granularity of the detector the DRN is able to account for the event-by-event variations in energy sharing between the electromagnetic and hadronic components. 
This is illustrated by the distribution of energy reconstructed in simulated events by the DRN (E,x,y,z) model and the WS method as a function of the fraction of incident energy carried by the $\pi^0$s for 50 GeV and 200 GeV pions as shown in Fig.~\ref{fig:drn-chi2-pi0frac}. 
The fraction of energy carried by $\pi^0$s is obtained directly from \GEANTfour.

\section{Expected performance using the full HGCAL geometry} \label{sec:fullhgcal}
\begin{figure}[h]
  \centering
  \includegraphics[width=0.45\linewidth]{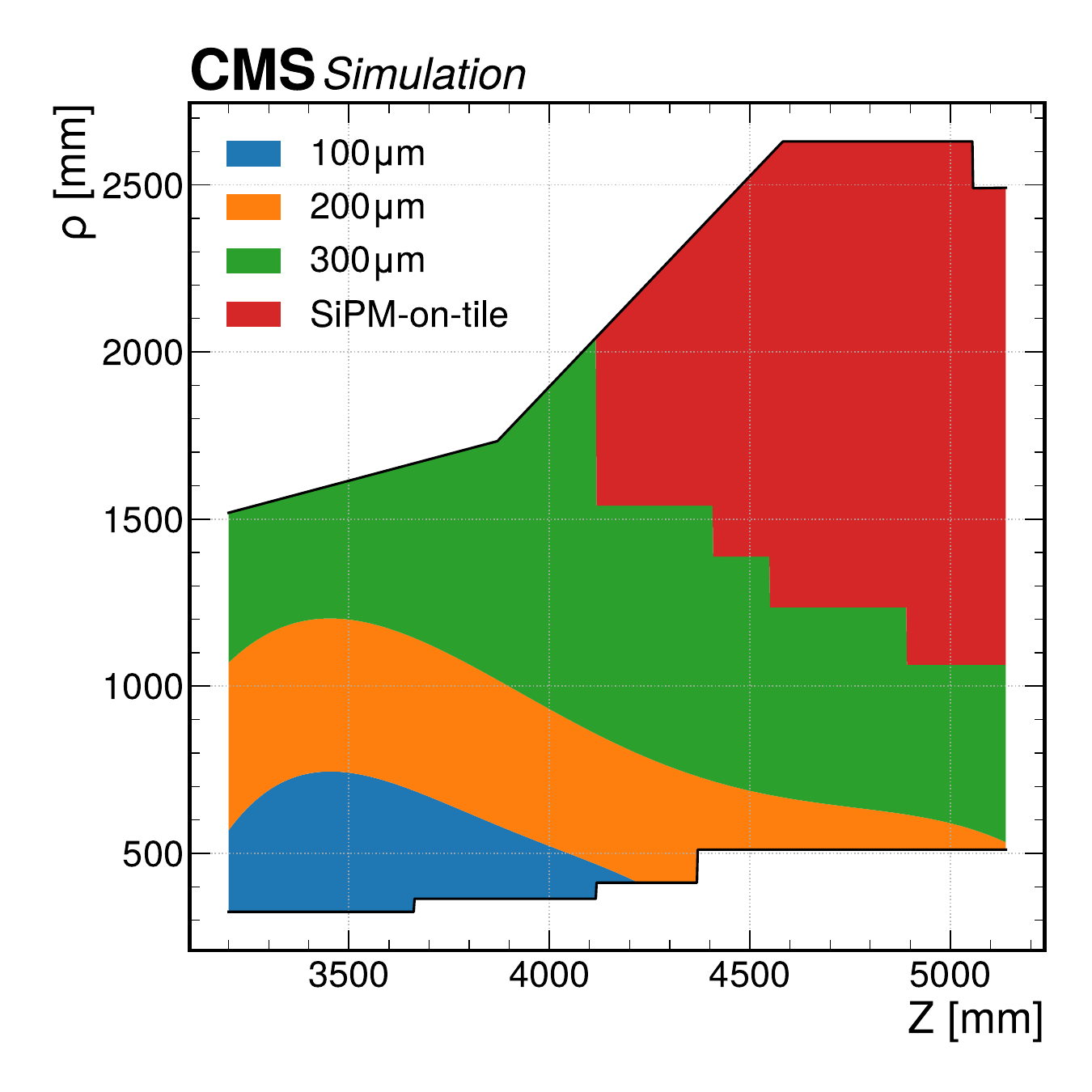}
  \caption{\label{fig:full-hgcal} Schematic of standalone geometry of expected HGCAL configuration.}
\end{figure}

Due to its limited size, energy losses from the prototype calorimeter were significant.
This will not be the case with the final HGCAL that will be installed in the CMS experiment.
To quantify the effect of potential leakage of energy and to validate the performance of the GNN-based algorithm in a realistic setup, we tested the model using a sample of pions generated with a standalone full-geometry setup simulated using \GEANTfour. 
The simulation did not include the upstream material that will be present in the final detector. 
It does, however, include a detector configuration of sensor modules with areas 0.5~$cm^2$ and 1~$cm^2$ and low density modules as shown in Fig.~\ref{fig:full-hgcal}, which reflects the final HGCAL configuration better than the one described in Ref.~\cite{bib.cms-hgcalphase2-tdr}. 
In this figure, $Z$ is the horizontal distance from the interaction point and $\rho$ is the radial distance from the beamline.

\begin{figure}[ht]
    \centering
    \includegraphics[width=0.45\textwidth]{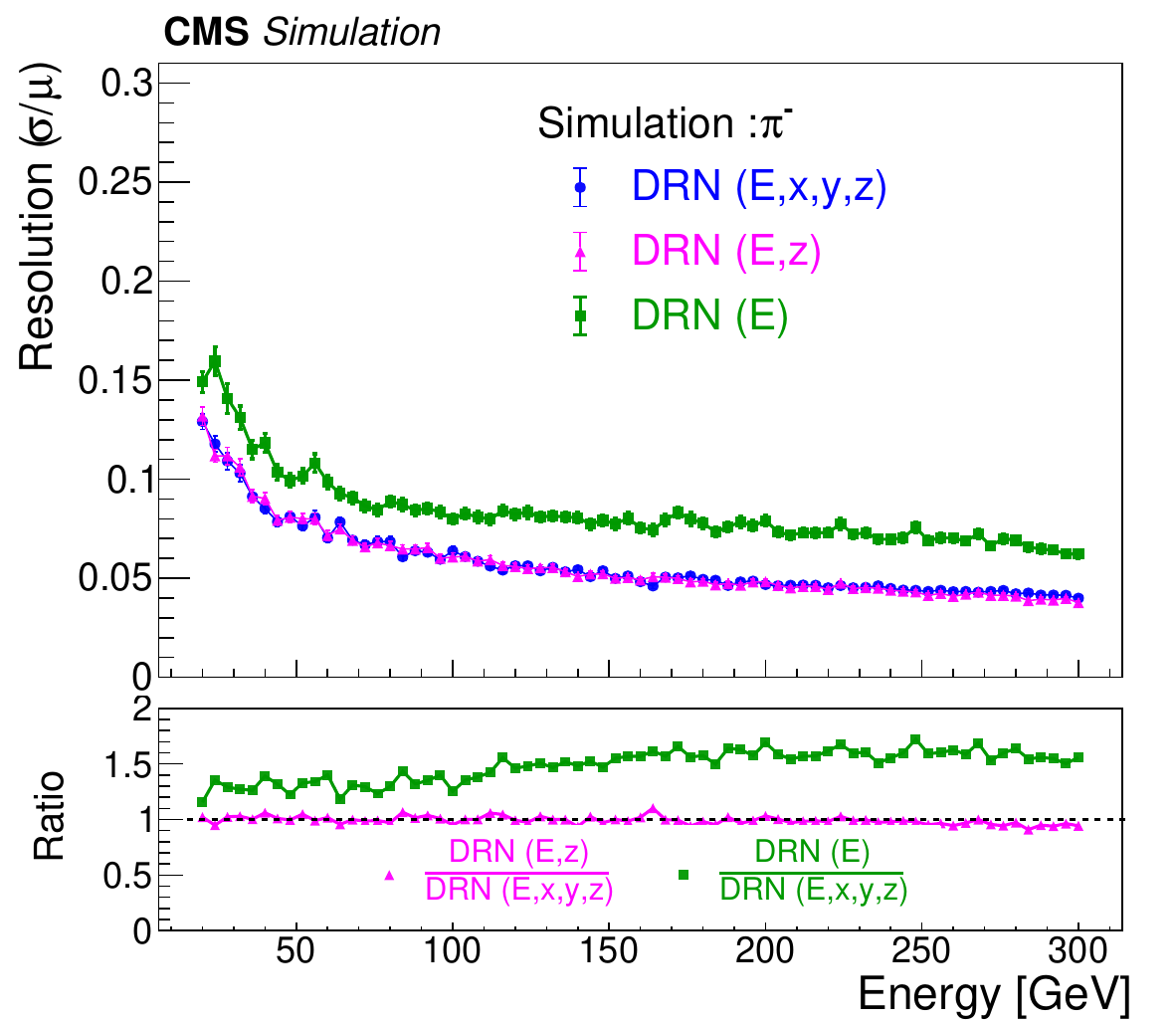}
    \includegraphics[width=0.45\textwidth]{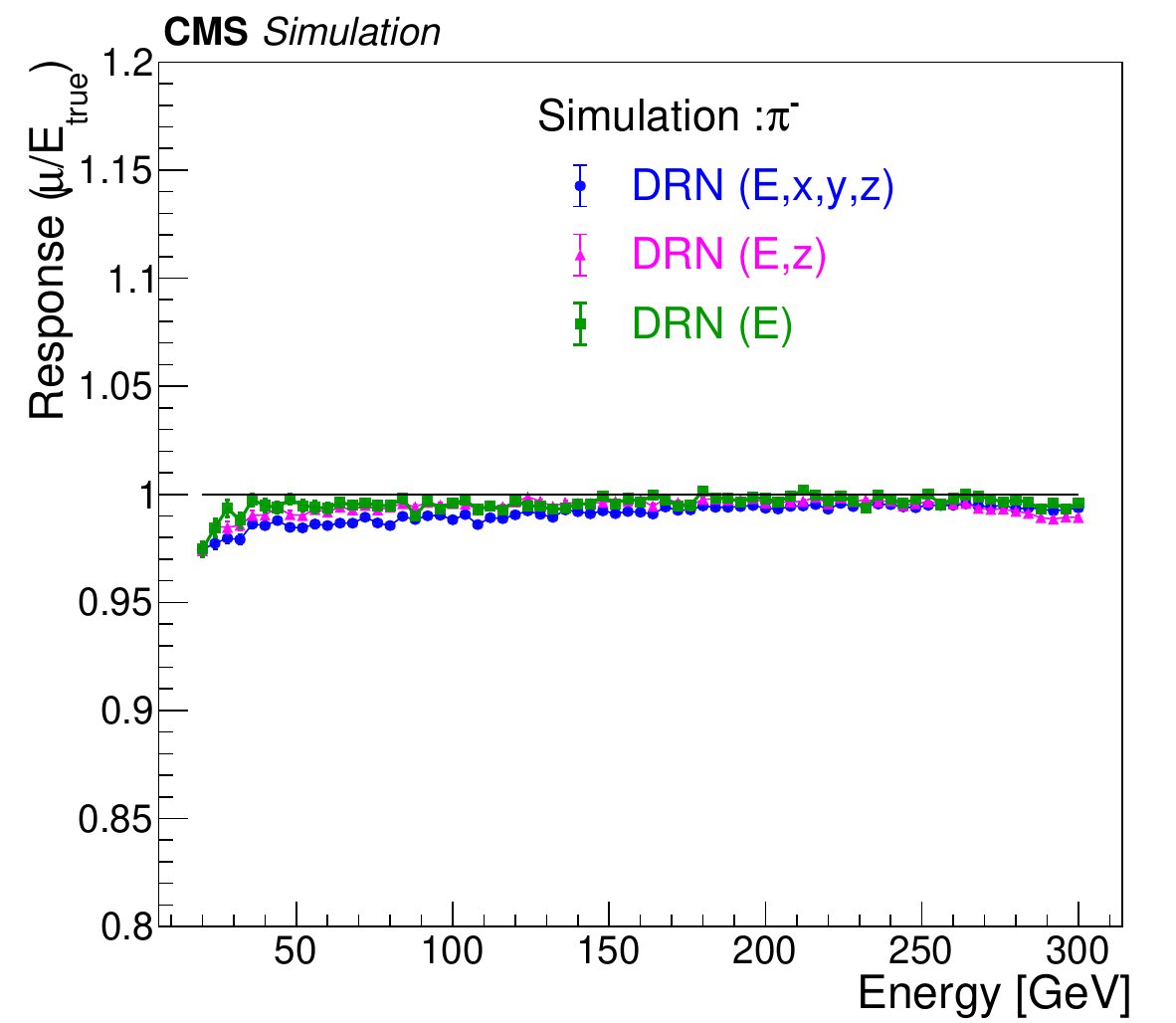}
    \caption{Resolution (left) and response (right) predicted by DRN(E), DRN(E,z) and DRN(E x y z) in the full HGCAL setup in simulation.}
    \label{fig:drn-fullhgcal}
\end{figure}

We simulated the pion showers entering at $\eta=2$ in this setup corresponding to $\rho$=800\,mm in Fig.~\ref{fig:drn-fullhgcal} and separately trained the model with the corresponding inputs.
As in the simulation of the beam test detector setup, the signal deposited in each detector channel is converted to the unit of number of MIPs by factors determined with simulated 200 GeV muons with the digitization and electronics noise superimposed on the signal. 
As before, rechits with energy deposits of less than 0.5 MIP are rejected. 
Three million charged pion events were simulated with the energies distributed uniformly from 10 to 350 GeV.
We use the rechits energies and the spatial coordinates to train the models DRN (E), DRN (E,z), and DRN (E,x,y,z). 
The resolutions and responses obtained with these three models are shown in Fig.~\ref{fig:drn-fullhgcal}.
We observed that adding information about the transverse coordinates of rechits has little impact on the predictions based on the DRN which is trained on input features comprising rechit energies and depth. 
This result can be understood as being due to the full shower containment in this extended geometry, unlike the smaller prototype where there was energy leakage.
The model incorporates additional information from the transverse distribution of rechits provided as $x$ and $y$ positions. 
For a fully contained shower, the model learns the true energy based on the highly granular distribution of energies of rechits and their depth in the detector.

\begin{figure}[ht]
    \centering
    \includegraphics[width=0.45\textwidth]{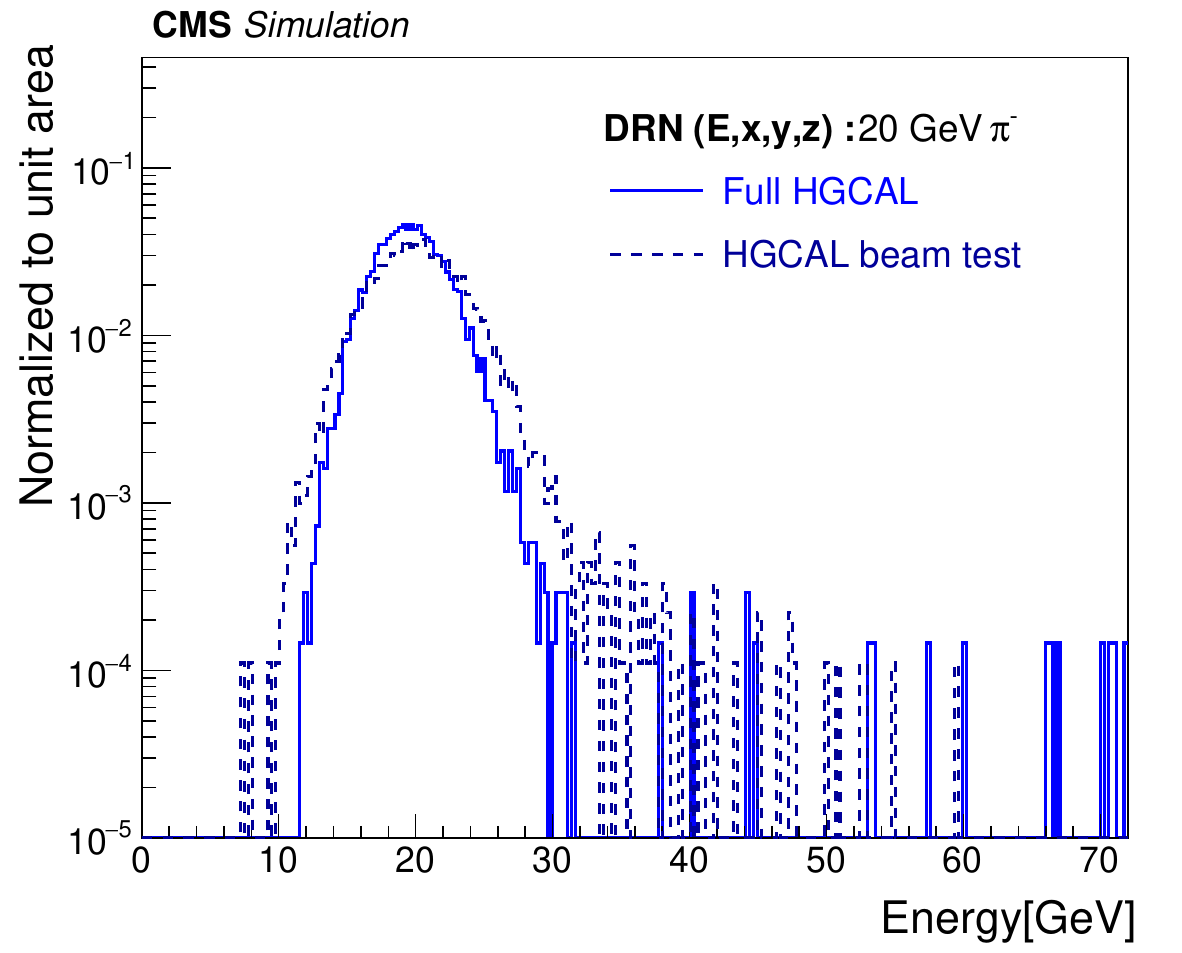}
    \includegraphics[width=0.45\textwidth]{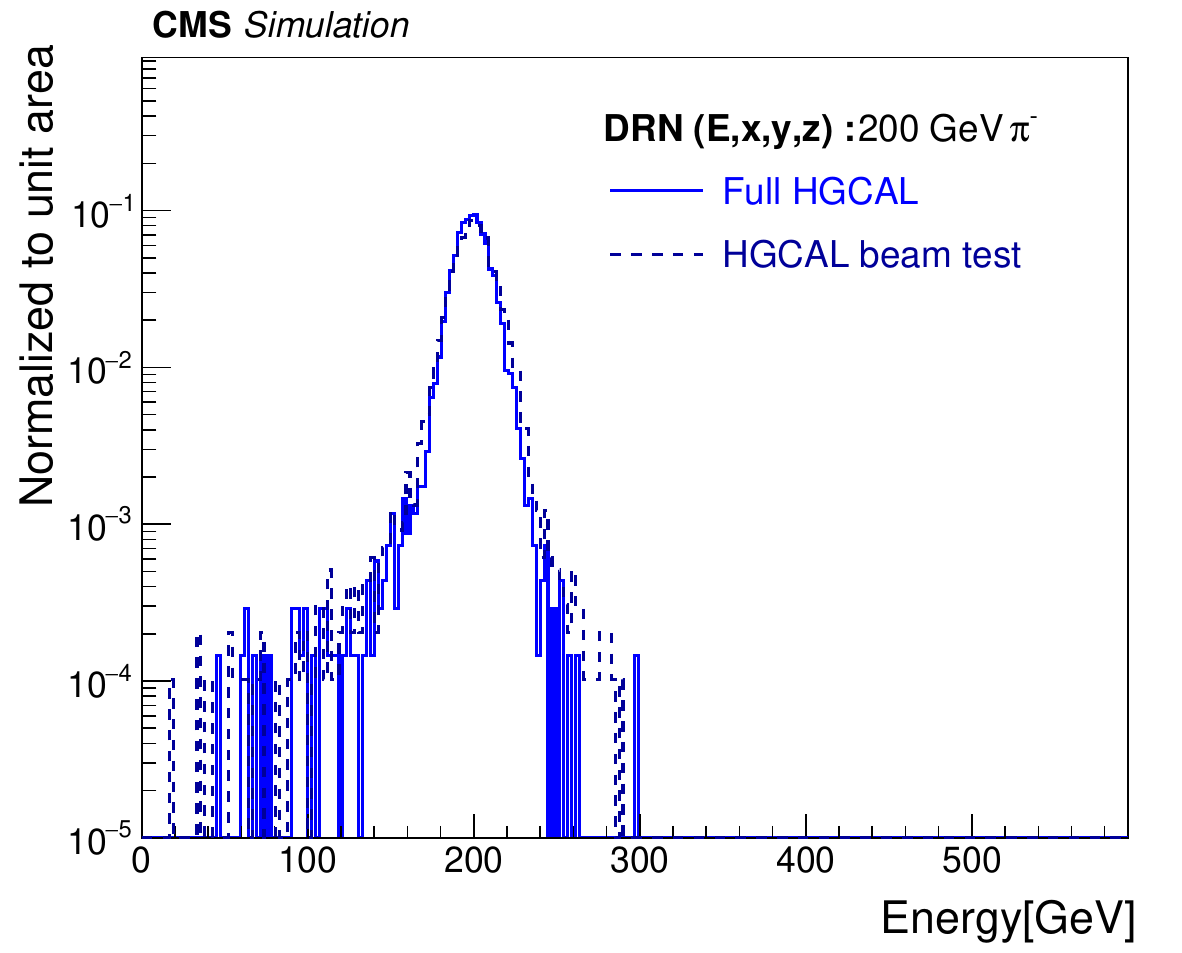}
    \caption{Distributions of energy reconstructed in full HGCAL and beam test setups using DRN(E,x,y,z) for pions with energies 20 GeV (left) and 200 GeV (right).}
    \label{fig:drn-fullhgcal-vs-tb-1d}
\end{figure}

\begin{figure}[ht]
    \centering
    \includegraphics[width=0.45\textwidth]{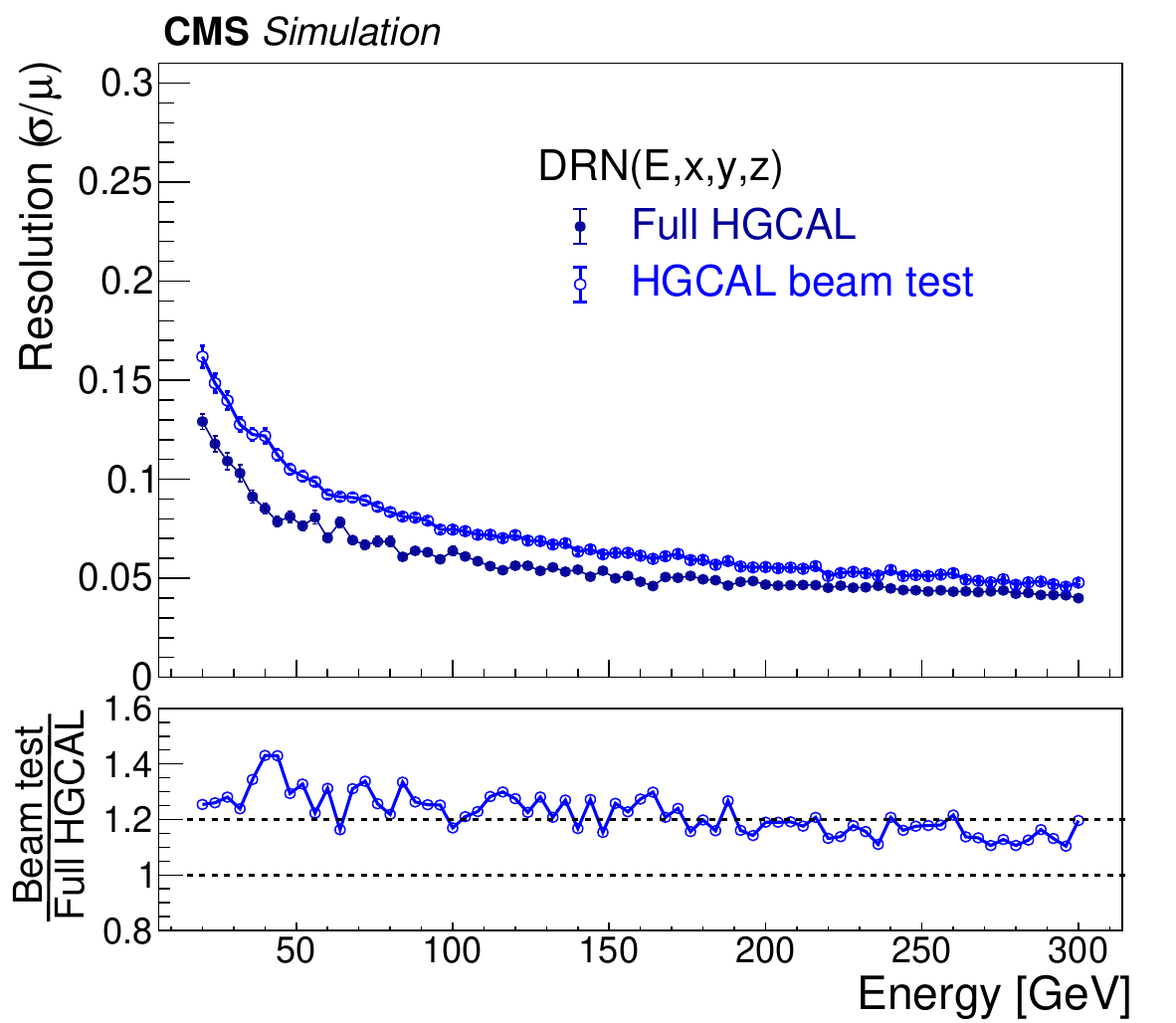}
    \includegraphics[width=0.45\textwidth]{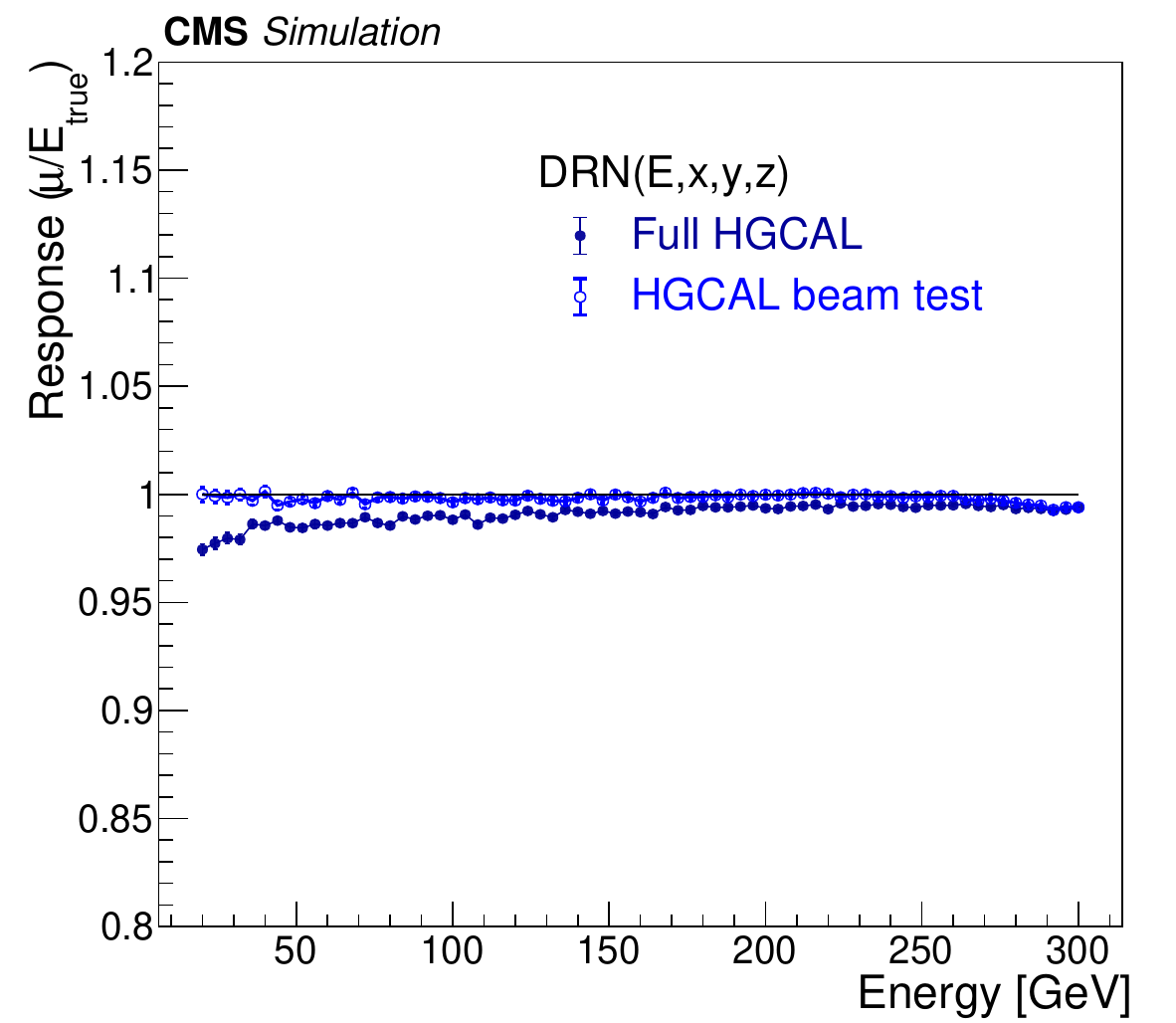}
    \caption{Comparison of energy resolution (left) and response (right) in full HGCAL and beam test setups using DRN(E,x,y,z).}
    \label{fig:drn-fullhgcal-vs-tb}
\end{figure}

Comparisons of the reconstructed energy distributions of the pions with 20 GeV and 200 GeV in the full HGCAL and beam test setups using DRN(E,x,y,z) are shown in Fig.~\ref{fig:drn-fullhgcal-vs-tb-1d}. The resolution and response as a function of incident pion energy of 20--300 GeV for the two setups are shown in Fig.~\ref{fig:drn-fullhgcal-vs-tb}.
We observe up to 30\% improvement in resolution predicted by DRN (E,x,y,z) at lower energies and up to 20\% at higher energies. 

\section{Summary} \label{sec:summary}
We have studied the performance of a dynamic reduction network (DRN) based on graphical neural networks (GNN) for the reconstruction of hadron showers from charged pions in a prototype of the CMS high granularity calorimeter (HGCAL) detector in a beam test conducted at CERN SPS in October 2018. 
We have found that the DRN improves by up to a factor of two the hadronic energy resolution compared to the conventional approach of summing up the energy deposited in various readout channels of the detector. 
We have also demonstrated that the use of transverse and longitudinal spatial information along with signal amplitudes makes precise intercalibration of detector cells across electromagnetic and hadronic sections unnecessary. 
Furthermore, the performance of the DRN model has been evaluated using a simulation of a full HGCAL geometry configuration, demonstrating overall improvement in the final resolution over conventional summation methods.
This is the first report evaluating the performance of a machine learning based energy reconstruction algorithm for hadron showers in the CMS HGCAL prototype using data collected in a beam test experiment. 

\acknowledgments{We thank the technical and administrative staffs at CERN and at other CMS institutes for their contributions to the success of the CMS effort. We acknowledge the enduring support provided by the following funding agencies and laboratories: BMBWF and FWF (Austria); CERN; CAS, MoST, and NSFC (China); MSES and CSF (Croatia); CEA, CNRS/IN2P3 and P2IO LabEx (ANR-10-LABX-0038) (France); SRNSF and GTU (Georgia); BMBF, DFG, and HGF (Germany); GSRT (Greece); DAE and DST (India); MES (Latvia); MOE and UM(Malaysia); MOS (Montenegro); PAEC (Pakistan); FCT (Portugal); NSTC (Taipei); ThEP Center, IPST, STAR, and NSTDA (Thailand); TUBITAK and TENMAK (Turkey); STFC (United Kingdom); and DOE (USA). The authors acknowledge the Minnesota Supercomputing Institute (MSI) at the University of Minnesota and the National Supercomputing Mission (NSM) at the Indian Institute of Science Education and Research, Pune for providing resources that contributed to the research results reported within this paper.}

\newpage
\bibliographystyle{JHEP}
\bibliography{main}

\end{document}